\documentclass[aps,prb,superscriptaddress,amsmath,showpacs,floatfix]{revtex4-2}
\usepackage{graphicx}
\usepackage{dcolumn}
\usepackage{bm}
\usepackage{subfig}
\usepackage{float}
\usepackage{color}
\usepackage[compat=1.1.0]{tikz-feynhand}

\begin{document}

\preprint{APS/123-QED}

\title{Application of Many-body Non-perturbative Theories to the Three-Dimensional Attractive Hubbard Model}


\author{Junnian Xiong}
\affiliation{
School of Physics, Peking University, Beijing 100871, China
} 

\author{Hui Li}
\affiliation{Institute for Advanced Study in Physics, Zhejiang University, Hangzhou 310058, China}

\author{Yingze Su}
\affiliation{
School of Physics, Peking University, Beijing 100871, China
} 

\author{Dingping Li}%
\email{lidp@pku.edu.cn}
\affiliation{
School of Physics, Peking University, Beijing 100871, China
}

\date{\today}

\begin{abstract}

The attractive Fermi-Hubbard model stands out as a simple model for studying the pairing and superconductivity of fermions on a lattice. 
In this article, we apply several many-body theories in the three-dimensional attractive Hubbard model. 
Specifically, we compare the results of various GW methods with DQMC simulations and observe that they provide reliable results in the weak to intermediate coupling regime. 
The critical exponents also agree well with the accurate results obtained from the 3D XY model.
In the superconducting phase, the post-GW method significantly improves the description of Green's functions and
density of states. Additionally, we propose a method to determine the temperature at which the pseudogap appears.

\end{abstract}

\maketitle


\section{\label{sec:intro}Introduction}

The Fermi-Hubbard model serves as a fundamental condensed matter model for strongly correlated fermionic systems \cite{auerbach_interacting_1994}. The attractive Hubbard model (AHM) has been used in various studies, including two-dimensional investigations of the Kosterlitz-Thouless (KT) transition \cite{2DDQMC,2DDQMC2}, and in the exploration of the positive-U Hubbard model through particle-hole transformations \cite{review2}. Primarily, the AHM provides a platform for studying the crossover from Bardeen-Cooper-Schrieffer (BCS) superconductivity \cite{BCS} to Bose-Einstein condensation (BEC) of local pairs \cite{narrow-band_systems,NSR,zwerger_bcs-bec_2012}. 
This crossover is important for both high-temperature superconductivity \cite{zwerger_bcs-bec_2012,PoHTS_1998,RN17} and 
cold-atom systems \cite{review1,cold_atom1,cold_atom2,cold_atom3,cold_atom4,UFG,ultracold_atom,Feshbach1}. While Fermi gas systems are also widely used to study this crossover, there are certain delicacies concerning the 
renormalization and discretization of continuum theories to lattice systems \cite{gase_to_lattice1,gase_to_lattice2}. 
In contrast, the AHM can be simulated using the 
Determinant Quantum Monte Carlo (DQMC) method \cite{3DDQMC,3DDQMC2}, which is free from the sign problem and provides precise results. 
Experimental studies on the AHM have been conducted in optical lattices \cite{UAE_of_AHM1,UAE_of_AHM2,UAE_of_AHM3,cold_atom4,review2}, 
providing valuable references for theoretical studies \cite{Experimental1,Brown_experiment}. 

In the 3D AHM, there are mainly two phases: the broken-symmetry phase below the superconducting critical 
temperature $T_c$ and the normal phase above it.
In the normal phase, apart from the BCS limit regime, 
preformed pairs exist up to a temperature, 
$T^*$ (higher than the superconducting critical temperature). 
When the temperature goes below the superconducting critical temperature 
$T_c$, coherence among the pairs is established \cite{review_Chen2,randeria_crossover_2014,review_Chen1}. As the system transitions from the BCS to the BEC regime, these two phenomena exhibit different behaviors. 

Various many-body methods,  
such as T-matrix approximation (TMA) \cite{review_Chen1,review_Chen2,review3}, dynamical mean-field theory (DMFT) \cite{DMFT1,DMFT2,DMFT3}, 
dynamical vertex approximation (D$\Gamma$A) \cite{DGA1,DGA2} and the two-particle self-consistent approach (TPSC) \cite{Tremblay,Tremblay_prepair,Koinov}, 
have been applied to study the phenomena in the BCS-BEC crossover. 
The TMA, developed over recent decades from the perturbative NSR theory \cite{NSR} to 
the more advanced self-consistent formulations by Haussmann\cite{Haussmann_selfcon}, 
has found wide application \cite{perturbative_tma1,GorKov_Popov,TMA1,tma3,self-consistent_tma1}. 
However, many of the methods are mainly used to study Fermi gases, where reliable numerical results obtained from Monte Carlo simulations are relatively scarce \cite{gase_to_lattice2,gas_MC}. The results in the 3D AHM are limited, especially in the discussion of pre-pairing and the relevant calculations in the superconducting phase.
Since DQMC provides accurate results in the 3D AHM, it is worthwhile to verify the effectiveness of different theoretical methods, and at the same time, it can serve as a reference for theories in Fermi gases.

In this paper, we apply many-body non-perturbative methods \cite{Hedin,general_GW,FLEX} in the 3D AHM and compare numerical results with DQMC simulations. 
Such methods include the GW \cite{LiHui} and HGW \cite{SunZhiPeng} approximations (commonly referred to as $G_0G$ in normal phase), 
which provides a many-body self-consistent description of the TMA. 
These methods have been applied to strongly correlated systems \cite{tma3,Haussmann_BCS,Parcollet}. 
The GW approximation is obtained from the first-order truncation of the Hedin equations \cite{Hedin,general_GW}, 
which contains the screened effect in its self-energy. 
The HGW method, based on the clustering properties of connected correlators, 
truncates high-order correlators in the Dyson-Schwinger equation and provides more accurate results in the intermediate-coupling regime \cite{SunZhiPeng}. 
Both methods have been used to calculate Green's functions.

To calculate the physical response functions, the covariant framework is proposed to preserve both the Fluctuation-Dissipation 
Theorem (FDT) and the Ward-Takahashi Identity (WTI) \cite{LiHui} 
in many-body approximation theories. 
This framework has been shown to significantly improve the accuracy of spin correlations for GW in the two-dimensional Hubbard model. 
Moreover, to address the unphysical effects caused by truncations in the GW approximation for correlated systems, the 
post-GW framework is proposed based on the relationship between the screened potential and the covariant response function. 
Ref. \cite{LiHui} shows that the post-GW method can describe the pseudogap, while the traditional GW method fails.  

We specifically apply many-body methods to the 3D AHM in regions that include the normal phase and the superconducting phase. 
In the normal phase, the GW method yields Green's functions, 
two-body correlation functions, and a transition temperature in the weak coupling regime 
that agrees well with DQMC results and reproduces the 3D XY model's critical exponents. 
The HGW method, on the other hand, provides accurate results in the intermediate coupling regime, particularly away from half-filling. 
In the superconducting phase, the post-GW method improves the description of the Green's function and the density of states. Additionally, we propose a method to determine the temperature at which the pseudogap appears.

This paper is organized as follows. In Sec.~\ref*{sec:formation}, 
we present the formalism of the GW and HGW approximations 
for Green's function, 
along with the covariant scheme for the GW approximation and post-GW corrections. 
In Sec.~\ref*{sec:result}, 
we first present the phase diagram results, 
followed by a comparison of Green's function 
and density of states with those obtained from DQMC. Finally, 
we calculate the two-body correlation function and associated critical exponents.
The conclusion and discussion are given in Sec.~\ref{sec:conclusion}.

\section{\label{sec:formation}FORMALISM}
\begin{figure*}
    \includegraphics[width=1\linewidth]{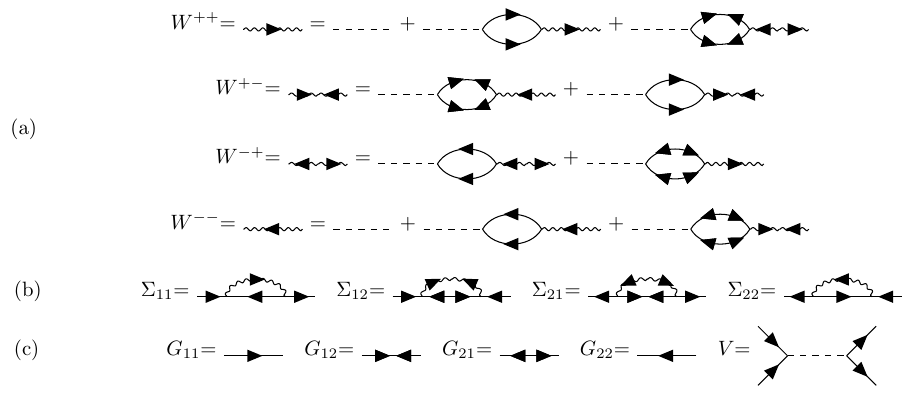}
    \caption{\label{fig:fenman1} 
    The Feynman diagrams of the GW equations. 
    (a) shows the Feynman diagram for the screened potential $W$. 
    If we introduce an auxiliary bosonic field $\phi$ using the Hubbard-Stratonovich 
    transformation to represent $W$, its definition can be written as: 
    $W^{++}=\langle\phi^*\phi\rangle$, $W^{+-}=\langle\phi\phi\rangle$, $W^{-+}=\langle\phi^*\phi^*\rangle$, 
    $W^{--}=\langle\phi\phi^*\rangle$. 
    (b) shows the self-energy diagram in the GW equations, and (c) depicts the Green's 
    function with anomalous components and the allowed four-fermion interaction in the equations.
    The Feynman diagram for the HGW method can be obtained by replacing one of the 
    dressed Green's functions in the loop diagrams with a Green's function that contains only the Hartree term.}
\end{figure*}

\subsection{\label{sec:DSE}Attractive Hubbard model and Dyson-Schwinger equation in Nambu space}
We consider the attractive Hubbard model with periodic boundary conditions on a cubic lattice 
\begin{equation}
    \hat H = -t\sum_{\langle ij\rangle}\sum_{\sigma=\uparrow,\downarrow}
    \hat c^\dagger_{i\sigma}\hat c_{j\sigma} - |U|\sum_{i}\hat n_{i\uparrow}\hat n_{i\downarrow}
    -\mu\sum_{i\sigma}\hat n_{i\sigma}.
\end{equation}
Here $\hat c^\dagger_{i\sigma}$ is the creation operator that creates an electron 
with spin $\sigma$ at lattice site $\textbf{i}$ and the corresponding $c_{i\sigma}$ 
is the annihilation operator. $\hat n_{i\sigma}=\hat c^\dagger_{i\sigma}c_{i\sigma}$ 
denotes the spin-resolved density operator. $t$ is the hopping amplitude, $U$ is the 
on-site interaction, and $\mu$ is the chemical potential. In this paper, we only consider 
the nearest-neighbor hopping and all energies are given in units of $t=1$ with the 
$\hbar = 1$.

To investigate the superconducting properties, 
one can rewrite the Hubbard Hamiltonian as 
\begin{equation}
    \hat {H} = -t\sum_{\langle i j\rangle\sigma}
    \hat c^\dagger_{i\sigma}\hat c_{j\sigma}
    -|U|\sum_{i}\hat \Delta^\dagger_{i}\hat \Delta_{i}-\mu\sum_{i \sigma}
    \hat c^\dagger_{i\sigma}\hat c_{i\sigma},\label{Cooper Hamiltonian}
\end{equation}
where $\hat \Delta^\dagger_{i}=\hat c^\dagger_{i\uparrow}
\hat c^\dagger_{i\downarrow}$. 
The Matsubara action for a fermionic system with a Cooper-Cooper-type interaction at finite temperature has the form
\begin{equation}
\begin{aligned}
    S[\psi,\psi^*]=&-\int\mathrm{d}(12)\sum_{\sigma}
    \psi^*_\sigma(1)T(1,2)\psi_\sigma(2)\\
    &+\int\mathrm{d}(12)V(1,2)\Delta^*(1)\Delta(2),\label{generalized action}
\end{aligned}
\end{equation}
where $\psi^*$, $\psi$ are Grassmannian fields and $\Delta^*(1)=\psi^*_{\uparrow}(1)\psi^*_{\downarrow}(1)$, 
$\Delta(1)=\psi_{\downarrow}(1)\psi_{\uparrow}(1)$ are the singlet pair-field operators. The label 
$(1)\equiv (\tau_1,\vec x_1)$ is the generalized coordinate, containing the Matsubara time 
$0\leq\tau_1\leq\beta$, where $\beta$ is the inverse temperature, and 
the space coordinate $\vec x_1$. The notation $\int\mathrm{d}(1)$ stands for integral over all space and time coordinates 
in the continuous system, and stands for summation in a discrete system.
The $T$ and $V$  
represent the kinetic term and the interaction potential, respectively.

In the superconducting case in Nambu space, 
we define the Nambu matrix Green's function \cite{Nambu} in ensemble average form 
\begin{equation}
    \underline{G} (1,2)
    =\left(\begin{array}{ll}
        \langle\psi^*_\uparrow(2)\psi_\uparrow(1)\rangle & \langle\psi_\downarrow(2)\psi_\uparrow(1)\rangle\\
        \langle\psi^*_\uparrow(2)\psi^*_\downarrow(1)\rangle & \langle\psi_\downarrow(2)\psi^*_\downarrow(1)\rangle
    \end{array}\right).\label{Nambu Green's function}
\end{equation}
Here $\langle \dots\rangle$ stands for $\frac{1}{Z}\int\mathrm{D}[\psi^*,\psi]\cdots e^{-S}$, where $Z=\int\mathrm{D}[\psi^*,\psi]e^{-S}$ 
is the grand partition function, and $D[\psi^*,\psi]$ defines the measure for 
path integration of fermion coherent states.

Consider adding external bosonic sources coupled to the Cooper pair operator:
\begin{equation}
\begin{aligned}
    S[\psi^*,\psi;J,J^*]=&S[\psi^*,\psi]-\int\mathrm{d}(1)J^{-}(1)\Delta(1)\\
    &-\int\mathrm{d}(1)J^{+}(1)\Delta^*(1).
\end{aligned}\label{general action}
\end{equation}
Then Green's function and its functional derivative are related through the Dyson-Schwinger equation
\begin{equation}
\begin{aligned}
    \underline{I}\delta(1,2)=&\int\mathrm{d}(3) \underline{H}^{-1}(1,3)\underline{G}(3,2)\\
    &-\int\mathrm{d}(3)V(1,3)\underline{\sigma}^{+}\frac{\delta \underline{G}(1,2)}{\delta J^{-}(3)}\\
    &-\int\mathrm{d}(3)V(1,3)\underline{\sigma}^{-}\frac{\delta \underline{G}(1,2)}{\delta J^{+}(3)},\label{Dyson-Schwinger}
\end{aligned}
\end{equation}
where $\underline{\sigma}$ matrices are defined by 
$\underline{\sigma}^{+} = \frac{1}{2}\left(\underline{\sigma}^{x}+\mathrm{i}\underline{\sigma}^{y}\right)$, 
$\underline{\sigma}^{-} = \frac{1}{2}\left(\underline{\sigma}^{x}-\mathrm{i}\underline{\sigma}^{y}\right)$. 
The underline $\underline{X}$ means matrix form in Nambu space, and
$\underline{I}$ is the identity matrix in Nambu space. 
Here $\delta(1,2)$ is the Dirac/Kronecker delta function for continuous/discrete systems. 
The Hartree-like propagator $H$ is defined by
\begin{equation}
    \underline{H}^{-1}(1,2)\equiv \underline{T}(1,2)+\underline{\sigma}^{+}\delta(1,2)v^{+}(1)
    +\underline{\sigma}^{-}\delta(1,2)v^{-}(1).\label{H format}
\end{equation}
Here the single-particle effective potentials are
\begin{subequations}
    \begin{align}
        v^{+}(1)=&J^{+}(1)-\int\mathrm{d}(3)V(1,3)\langle\Delta(3)\rangle,\\
        v^{-}(1)=&J^{-}(1)-\int\mathrm{d}(3)V(1,3)\langle\Delta^{*}(3)\rangle,
    \end{align}\label{v format}
\end{subequations}
and the kinetic term is
\begin{equation}
    \underline{T}(1,2)=\left(\begin{array}{ll}
        T(1,2) & 0\\
        0 & -T(2,1)
    \end{array}\right).
\end{equation}

\subsection{\label{sec:HGWandGGW}GW and HGW approximation}

According to Hedin's vertex approximation, we can obtain
\begin{equation}
    \frac{\delta \underline{G}^{-1}(1,2)}{\delta v^{a}(3)}\approx
    \frac{\delta \underline{H}^{-1}(1,2)}{\delta v^{a}(3)}=\underline{\sigma}^{a}\delta(1,2)\delta(1,3).\label{GGW truncation}
\end{equation}
Here $a$ stands for $+$ or $-$. Apply this truncation to the Eq.~(\ref*{Dyson-Schwinger}), we can get the GW equations in 
Nambu space
\begin{subequations}
    \begin{align}
        \underline{G}^{-1}(1,2)=&\underline{H}^{-1}(1,2)-\underline{\Sigma}_{\mathrm{G}}(1,2),\\
        \underline{\Sigma}_{\mathrm{G}}(1,2)=&-\sum_{a,b=\pm}\underline{\sigma}^{a}
        \underline{G}(1,2)\underline{\sigma}^{b}W^{b\bar a}_{\mathrm{G}}(2,1),\\
        W^{ab-1}_{\mathrm{G}}(1,2)=&V^{-1}(1,2)I^{ab}-\Pi^{ab}_{\mathrm{G}}(1,2),\\
        \Pi^{ab}_{\mathrm{G}}(1,2)=&\mathrm{Tr}\left[\underline{\sigma}^{\bar a}\underline{G}(1,2)
        \underline{\sigma}^{b}\underline{G}(2,1)\right].
    \end{align}\label{GGW equation}
\end{subequations}
Here, $\Pi$ is the polarization function, $W$ is the screened potential, $\Sigma$ is the GW self-energy, and $\bar a = -a$. 
The trace in the equations is performed within Nambu space. 
These equations can be solved self-consistently. 

In addition to approximating the vertex, we can also obtain a new method by truncating the correlation function. Taking the functional derivative of Eq.~(\ref*{Dyson-Schwinger}) with respect to the source $J^a$ yields 
\begin{equation}
\begin{aligned}
    0=&\int\mathrm{d}(3)\frac{\delta \underline{H}^{-1}(1,3)}{\delta J^{a}(4)}\underline{G}(3,2)\\
    &+\int\mathrm{d}(3)\underline{H}^{-1}(1,3)\frac{\delta \underline{G}(3,2)}{\delta J^{a}(4)}\\
    &-\int\mathrm{d}(3)V(1,3)\underline{\sigma}^{+}\frac{\delta^2 \underline{G}(1,2)}{\delta J^{a}(3)\delta J^{-}(4)}\\
    &-\int\mathrm{d}(3)V(1,3)\underline{\sigma}^{-}\frac{\delta^2 \underline{G}(1,2)}{\delta J^{a}(3)\delta J^{+}(4)}.\label{HGWorigin}
\end{aligned}
\end{equation}
This equation relates the one-body correlator $\underline{G}$, the two-body correlator $\frac{\delta \underline{G}}{\delta J^a}$ 
and the three-body correlator $\frac{\delta^2 \underline{G}}{\delta J^a\delta J^b}$. By continuing this functional 
derivative process, one can obtain a hierarchy relation of different order correlators. 

In order to implement numerical calculations, proper truncation is necessary. 
The simplest truncation is to neglect the term 
$\delta G/\delta J$ in the first-order Dyson-Schwinger equation Eq.~(\ref*{Dyson-Schwinger}), 
which implies $G=H$, i.e., the Hartree approximation. A more precise approach 
is to truncate $\delta^2G/\delta J^2$ in the Eq.~(\ref*{HGWorigin}), which is based on the clustering property \cite{SunZhiPeng}. With this truncation, we obtain the HGW equation (commonly known as $G_0G$ in normal phase):
\begin{subequations}
    \begin{align}
        \underline{G}^{-1}(1,2)=&\underline{H}^{-1}(1,2)-\underline{\Sigma}_{\mathrm{H}}(1,2),\\
        \underline{\Sigma}_{\mathrm{H}}(1,2)=&-\sum_{a,b=\pm}\underline{\sigma}^{a}
        \underline{H}(1,2)\underline{\sigma}^{b}W^{b\bar a}_{\mathrm{H}}(2,1),\\
        W^{ab-1}_{\mathrm{H}}(1,2)=&V^{-1}(1,2)I^{ab}-\Pi^{ab}_{\mathrm{H}}(1,2),\\
        \Pi^{ab}_{\mathrm{H}}(1,2)=&\mathrm{Tr}\left[\underline{\sigma}^{\bar a}\underline{H}(1,2)
        \underline{\sigma}^{b}\underline{G}(2,1)\right].
    \end{align}\label{HGW equation}
\end{subequations}

Essentially, the GW equations and the HGW equations are based on different approximations, but they share a similar equation form. 
The HGW method can be derived by replacing some of Green's functions in the GW equations 
with Green's functions that only include the Hartree term. For later reference, 
Fig.~\ref*{fig:fenman1} provides the diagrammatic representation of the GW equations. 
The detailed derivation process can be found in the Appendix~\ref*{sec:detailsofGW}.
These formulations will be used to calculate Green's 
function and the density. Next, we will address the problem of the two-body 
correlation functions.

\subsection{\label{sec:covariance}Covariant scheme}
\begin{figure*}
    \includegraphics[width=1\linewidth]{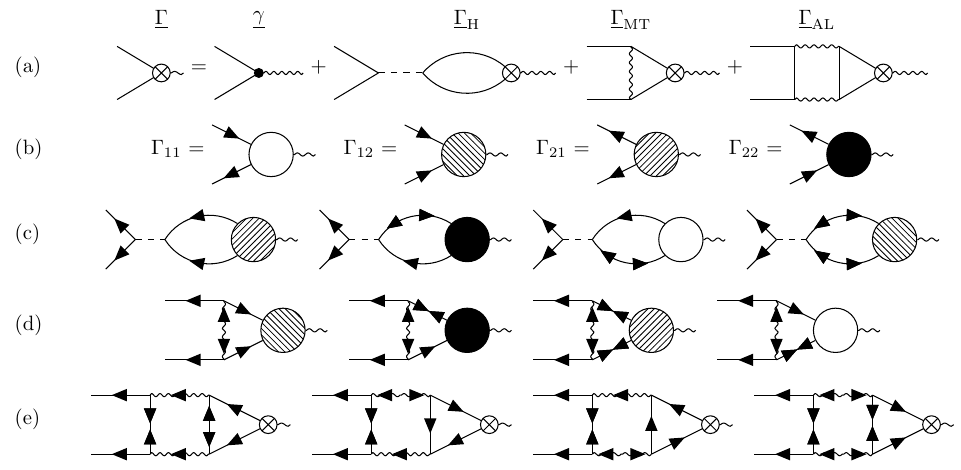}
    \caption{\label{fig:fenman2} 
    The Feynman diagrams for the covariant scheme. 
    (a) shows all possible topologies for the complete vertex. 
    (b) displays all possible vertices and the corresponding directions for the external legs. 
    (c) shows the Feynman diagram for $\Gamma_{\mathrm{H}}$ with a specific external leg direction. 
    (d) shows the Feynman diagram for $\Gamma_{\mathrm{MT}}$ with a specific external leg direction. 
    (e) shows the Feynman diagram for $\Gamma_{\mathrm{AL}}$ with a specific external leg direction. 
    There are four possible external leg directions in total. Due to the large number of complete 
    Feynman diagrams, only one is shown here; the others can be obtained by applying the 
    corresponding vertex rules.
    For the HGW method, it is necessary to replace 
    the vertex in $\Gamma_{\mathrm{MT}}$ with $\gamma + \Gamma_{\mathrm{H}}$, 
    and to replace the dressed Green's function in the $\Gamma_{\mathrm{AL}}$ 
    ring diagram with a Green's function that contains only the Hartree term. 
    The explicit expressions can be found in the Appendix~\ref*{apd:covariance}}
\end{figure*}
According to the FDT, the two-body correlation functions should be defined as the response of 
the physical quantity in the presence of an external potential, which we refer to as the covariant 
scheme \cite{LiHui}.

We consider the general two-body correlation function
$\chi_{XY}(1,2)=\langle X(1)Y(2)\rangle_c$, where $\langle \cdot\rangle_c$ denotes the connected correlation function, and $X$, $Y$ are local binary operators
and take the form $X(1)=\int\mathrm{d}(23)\psi^m(2)K^{mn}_{X}(1,2,3)\psi^{n}(3)$. By adding an 
external local source $\phi(1)$ which couples the local operator $Y(1)$, 
the perturbed action becomes
\begin{equation}
    S[\psi^*,\psi;\phi]=S[\psi^*,\psi] - \int\mathrm{d}(1)\phi(1)Y(1).
\end{equation}
The additional term can be regarded as a variation of the $T$ term
\begin{equation}
    \underline{T}(1,2;\phi) = \underline{T}(1,2) + \int\mathrm{d}(3)\phi(3)K^{mn}_{Y}(1,2,3).
\end{equation}
Then the two-body correlation function can be expressed as
$\chi_{XY}(1,2)=\frac{\delta \langle X(1)\rangle}{\delta \phi(2)}$. 
For s-wave pair correlation 
$K^{ab}_{\Delta}(1,2,3)=\delta(1,2)\delta(1,3)\delta^{\cdot\cdot}_{\downarrow\uparrow}$. 
Note that 
\begin{equation}
    \langle X(1)\rangle=\int\mathrm{d}(23) K^{mn}_{X}(1,2,3)G^{nm}(3,2).
\end{equation}
Then the two-body 
correlation function can be calculated through the equation
\begin{equation}
\begin{aligned}
    \chi_{XY}(1,2)=&-\int\mathrm{d}(3456)\mathrm{Tr}[\underline{K}_X(1,3,4)\underline{G}(3,5)\\
    &\underline{\Gamma}_\phi(5,6,2)\underline{G}(6,4)],\label{Covariant chi}
\end{aligned}
\end{equation}
where $\underline{\Gamma}_\phi$ is the vertex function, and can be decomposed as 
\begin{equation}
\begin{aligned}
    \underline{\Gamma}_\phi(1,2,3)\equiv&\frac{\delta \underline{G}^{-1}(1,2)}{\delta \phi(2)}\\
    =&\underline{\gamma}_\phi(1,2,3)+\underline{\Gamma}_{\phi,\mathrm{H}}(1,2,3)\\
    &+\underline{\Gamma}_{\phi,\mathrm{MT}}(1,2,3)+\underline{\Gamma}_{\phi,\mathrm{AL}}(1,2,3).
\end{aligned}\label{covariance}
\end{equation}
Among them, $\underline{\Gamma}_{\phi,\mathrm{MT}}$ represents the Maki-Thompson-like (MT) vertex, 
and $\underline{\Gamma}_{\phi,\mathrm{AL}}$ represents the Aslamazov-Larkin-like (AL) vertices. 
For simplicity of expression, we introduce the notation $\underline{\Lambda}_{\phi}(1,2,3)=\frac{\delta \underline{G}(1,2)}{\delta \phi(3)}$, 
$\underline{\Lambda}_{\phi,\mathrm{H}}(1,2,3)=\frac{\delta \underline{H}(1,2)}{\delta \phi(3)}$ and
$\underline{\Lambda}_{\phi,\mathrm{W}}(1,2,3)=\frac{\delta \underline{W}(1,2)}{\delta \phi(3)}$. 
The specific expressions for different vertices are as follows:
\begin{subequations}
    \begin{align}
        &\underline{\gamma}_\phi(1,2,3)
        =\underline{K}_{X}(1,2,3),\\
&\begin{aligned}
    &\underline{\Gamma}_{\phi,\mathrm{H}}(1,2,3)\\
    =&-\delta(1,2)\sum_{a}\int\mathrm{d}(4)V(1,4)\underline{\sigma}^{a}\mathrm{Tr}
    \left[\underline{\sigma}^{\bar a}
    \underline{\Lambda}_{\phi}(4,4,3)\right],
\end{aligned}\\
        &\underline{\Gamma}_{\phi,\mathrm{MT}}(1,2,3) = \sum_{a,b}\underline{\sigma}^{a}
        \underline{\Lambda}_{\phi}(1,2,3)\underline{\sigma}^{b}W^{b\bar a}(2,1),\\
        &\underline{\Gamma}_{\phi,\mathrm{AL}}(1,2,3) = \sum_{a,b}\underline{\sigma}^{a}
        \underline{G}(1,2)\underline{\sigma}^{b}\Lambda_{\phi,\mathrm{W}}^{b\bar a}(2,1,3).
    \end{align}\label{Covariant vertex equation}
\end{subequations}

The details for the derivation are presented in Appendix~\ref*{apd:covariance}. 
The Feynman diagrams are referenced in Fig.~\ref*{fig:fenman2}, 
with the vertex arrow rules outlined in Fig.~\ref*{fig:fenman1}. 
Fig.~\ref*{fig:fenman2} only shows part of the Feynman diagrams, 
as the complete vertex Feynman diagram involves four possible configurations of external legs, which leads to numerous diagrams that satisfy the rules. 
For brevity, we select one as an example. For the HGW method, 
a part of the Green's function in the diagram should be replaced by the 
Green's function containing only the Hartree term, 
and the vertex in $\Gamma_{\mathrm{MT}}$ should be replaced by $\gamma + \Gamma_{\mathrm{H}}$. 
The detailed form can be found in the Appendix~\ref*{apd:covariance}. 
Notably, due to symmetry protection, the $\Gamma_{MT}$ and $\Gamma_{AL}$ vertices are absent when calculating the pairing correlation in the normal phase (see Fig.~\ref*{fig:fenman2}). As a result, the covariance method reduces to the RPA method.
However, in the superconducting phase, using only the RPA to calculate correlation functions is inconsistent with the GW and HGW equations. 
Only the covariance scheme satisfies both the FDT and the WTI \cite{LiHui}. 

\subsection{\label{sec:post-GW}Post-GW}
The motivation for the post-GW method arises from the violation of intrinsic 
relationships for higher-order correlation functions in the original theories 
due to truncation, such as the relation between the screened potential and the charge/pair correlation. 
According to Hedin's equation (Eq.~\ref*{eq:Hedin}), the rigorous relationship between the screened potential and the correlation function, i.e., the functional derivative of Green's function with respect to the external source, can be written as
\begin{equation}
\begin{aligned}
        W^{ab}(1,2) =& V(1,2)I^{ab} \\
        &+ \int\mathrm{d}(34)V(1,4)\mathrm{Tr}\left[\sigma^{\bar a}
    \frac{\delta \underline{G}(4,4)}{\delta J^{b}(3)}\right]V(2,3).
\end{aligned}
\end{equation}
In the GW theory, this relation is violated due to the vertex approximation, as shown in Eq.~(\ref*{GGW truncation}). 
As discussed earlier, we introduce the covariant theory to obtain physical correlation functions, which preserve the FDT and the WTI. 
In our post-GW approach, we replace the screened potential in the GW Green's 
function with the physical potential, which is determined by the covariant response function \cite{postgw}: 
\begin{equation}
    W_{\mathrm{post}}^{ab}(1,2) = V(1,2)I^{ab} + \int\mathrm{d}(34)V(1,4)\chi_{\mathrm{cov}}^{ab}(4,3)V(2,3).\label{post equation}
\end{equation}
This, to some degree, restores the relationship between the screened potential $W$ and the response of the Green's function to the external source. 
Based on the newly obtained screened potential 
$W$, we then compute the updated Green's function. The new Green's function is given by:
\begin{equation}
\begin{aligned}
    \underline{G}^{-1}_{\mathrm{post}}(1,2)=&\underline{H}^{-1}(1,2)-\underline{\Sigma}_{\mathrm{post}}(1,2),\\
    \underline{\Sigma}_{\mathrm{post}}(1,2)=&-\sum_{a,b=\pm}\underline{\sigma}^{a}
    \underline{G}(1,2)\underline{\sigma}^{b}W^{b\bar a}_{\mathrm{post}}(2,1).\\
\end{aligned}
\end{equation}
As mentioned in Sec.~\ref*{sec:covariance}, when the interaction is decomposed into Cooper pairs, 
the pairing correlation function calculated using the GW method with RPA in the normal phase is consistent with the covariance. Therefore, in the normal phase, the GW method in this work does not require a post-correction. In the superconducting phase, however, the broken symmetry leads to more complex correlation behaviors, thus the post-correction is needed.

\section{\label{sec:result}NUMERICAL RESULTS FOR THREE-DIMENSIONAL ATTRACTIVE HUBBARD MODEL}

This section evaluates the GW and HGW approximations in the three-dimensional attractive Hubbard model. Notably, the DQMC method does not exhibit a sign problem in this model, which serves as a benchmark for assessing the accuracy of our approximation. We employed the Julia program provided by Ref.~\cite{SmoQyDQMC} to obtain DQMC results. 

\subsection{\label{sec:phase}Phase diagram}

\begin{figure}
    \includegraphics[width=1\linewidth]{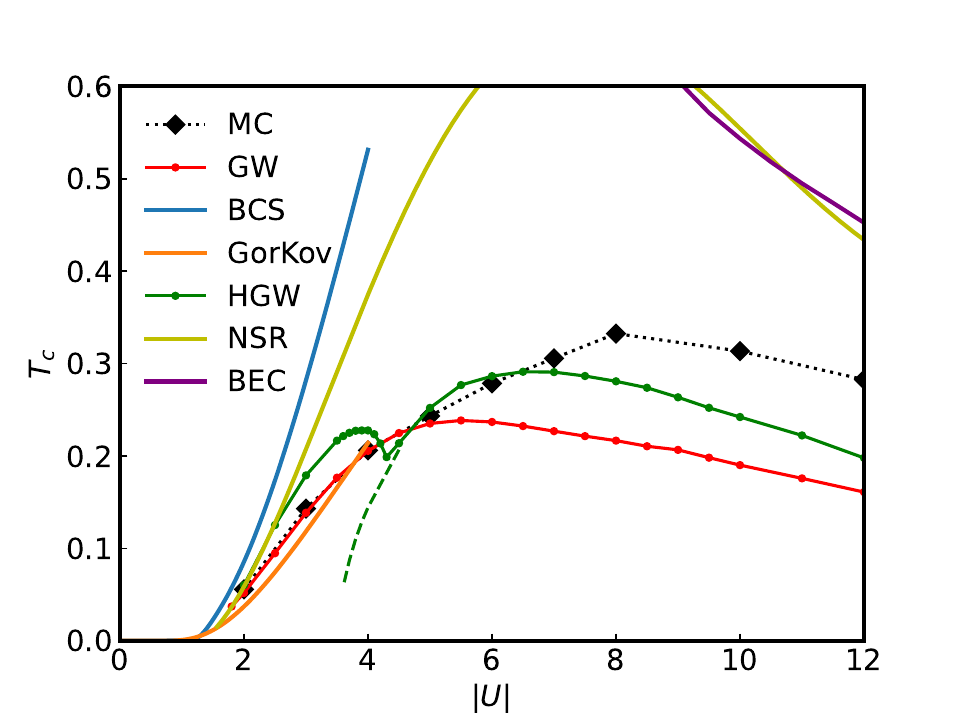}
\caption{\label{fig:tcvu} The plot shows the variation of the critical temperature with the coupling strength $U$ at a filling of $n = 0.5$. 
The phase transition temperatures obtained by various methods are displayed in the figure. 
These include DQMC results (black diamond line), BCS mean-field theory (blue line), 
Gorkov method (orange line) \cite{GorKov}, NSR method (yellow line) \cite{NSR}, BEC limit (green line), 
GW method (red dotted line), and HGW method (green dotted line). 
Among them, the green dashed line shows the critical temperature calculated by the HGW method with $W$ as the correlation function, which is done after the covariance method diverges.
}
\end{figure}

The phase transition temperature results obtained by various methods are compared with DQMC in Fig.~\ref*{fig:tcvu}. 
The DQMC results are based on the method outlined by Ref.~\cite{3DDQMC}. 
We further validated their results through repeated calculations, 
by controlling the density error to be within 0.005. 
Based on our recent computations, 
we added several additional data points, leading to the results presented here. 
In the small $|U|$ regime, the mean-field BCS theory yields $ T_c \propto \exp(1/|U|) $, 
which matches the behavior observed in DQMC. 
In the large $|U|$ limit, DQMC results follow the 3D-BEC formula, which yields $ T_c \propto 1/|U|$. 
While in the crossover region, 
we observe a smooth interpolation between the BCS and BEC regimes, with a peak around $U\approx8$.

To calculate the phase transition temperature curve for the GW and HGW methods, 
we need to compute the s-wave pair correlation function, which is defined as follows
\begin{equation}
    \begin{aligned}
        P_s(1,2)=&\langle \Delta^*(1)\Delta(2)\rangle\\ =& \langle \Delta^*(1)\Delta(2)\rangle_c 
        +\langle \Delta^*(1)\rangle\langle\Delta(2)\rangle.
\end{aligned}
\end{equation}
Here we consider a system with translational invariance, where $P_s(1, 2)=P_s(1 - 2)$. In the normal phase, the divergence of the s-wave pair correlation of Cooper pairs at 
$(i\omega_n,\vec k) = 0$ corresponds to the coherence between Cooper pairs.  
However, for the Hubbard model in a finite-size lattice system, the exact numerical value of the correlation function does not diverge, as confirmed by DQMC simulations \cite{3DDQMC}.

For the GW method, we determine the phase transition temperature in the thermodynamic 
limit by analyzing finite-size scaling behavior. First, we calculate the pairing correlation function using the covariance method in real space, from which we derive the correlation length. 
As the finite-size system approaches criticality, 
the correlation length satisfies \cite{Privman1990FiniteSS}:
\begin{equation}
\xi(T_c) \propto L,\label{eq:xi}
\end{equation}
which implies that the system exhibits long-range order. We identify this temperature as the critical temperature for a system of this size. Finally, the phase transition temperature in the thermodynamic limit is 
obtained through finite-size scaling \cite{3DDQMC,3DXY_book}:
\begin{equation}
    T_c(\infty)=T_c(N)+\mathcal{O}(1/\sqrt{N}).\label{tcfss}
\end{equation}
For detailed steps, please refer to Appendix~\ref*{sec:detailofTc}.

The situation is somewhat different for the HGW method. 
In the HGW framework, when the interaction strength $U$ is small, 
the results of covariance calculations diverge as the temperature gradually decreases, 
regardless of the system size.
In contrast, for larger values of $U$, the results of covariance calculations remain finite 
and do not show a significant increase as the temperature decreases. Meanwhile, $W$ in the HGW framework tends to increase with decreasing temperature, leading to instability in the normal phase solution as the temperature continues to drop.  
Therefore, in the HGW equation, we use $W$ as the s-wave pair correlation function result and apply the same method as the GW-covariance to determine the phase transition temperature in the large-$U$ regime. 
For small $U$, we similarly adopt the covariance method, using the temperature at which the correlation 
function diverges as the phase transition temperature. 
However, this approach leads to discontinuities in the HGW phase diagram, as illustrated in Fig.~\ref*{fig:tcvu}. 
On the other hand, if $W$ continues to be used as a criterion for phase transition determination, 
the result, as shown by the green dashed line in Fig.~\ref{fig:tcvu}, rapidly approaches zero.

\begin{figure*}[]
    \includegraphics[width=1.\linewidth]{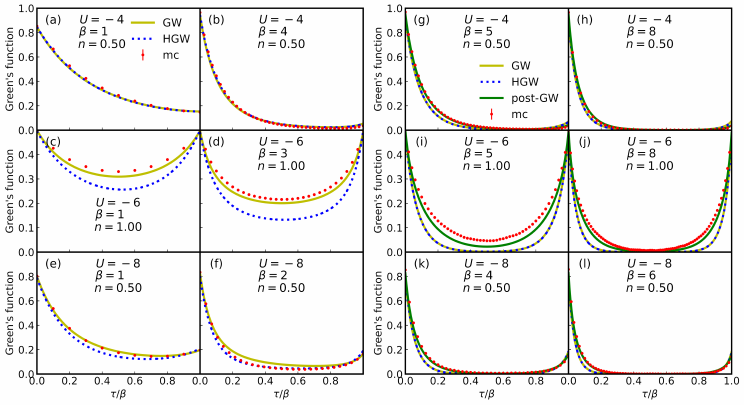}
    \caption{\label{fig:gtau} 
    Comparison of Green's function results on the Matsubara time 
    axis at the nodal point 
$k=(\pi/2,\pi/2,\pi/2)$ for the $8 \times 8 \times 8$ lattice under different parameters. The left part is the Green's function in the normal phase, while the right part is the Green's function in the superconducting phase.
    The results include DQMC (red dots), HGW method (black line), and 
    GW method (purple line). }
    \end{figure*}

The GW method provides results 
that are closer to the DQMC results when the absolute value of the attractive interaction $U$ is approximately below 5. While the HGW approximation is closer to the DQMC results compared to the GW approximation 
in regions for larger values of $|U|$. 
The GW method achieves its maximum critical temperature at $|U|\approx5$, 
while the HGW method attains its maximum critical temperature at $|U|\approx7$. 
Based on the two-body scattering formula, 
we find that when $|U|\approx7.915$, $1/k_F a=0$ \cite{Critical_Temperature_Fermions}, which corresponds to the unitarity region.
These results can be understood as follows: the HGW approximation truncates three-body connected correlators, which have good clustering properties at strong coupling $|U|$. However, at small $|U|$, these correlators may become nonlocal, making truncation less suitable. 
Both the GW and HGW methods capture the BCS-BEC crossover behavior but show significant deviations from DQMC results in the crossover region near the BEC limit.

In this article, GW and HGW methods focus exclusively on the particle-particle channel. 
For Fermi gas calculations, the Gorkov method \cite{GorKov,GorKov2} is used, 
which incorporates the particle-hole channel to correct the electronic screened potential. 
In the BCS limit, the Gorkov method averages the particle-hole channel to modify the coupling strength \cite{GorKov,GorKov2,GorKov_Popov}, 
yielding the following equation:
\begin{equation}
    0=\frac{1}{U}+\langle\chi_{ph}\rangle +\chi_{pp}(i\omega_n=0,\vec k=0)
\end{equation}
Here, $\chi_{ph}$ is calculated at the Lindhard level. 
The result is represented by the orange line in Fig.~\ref*{fig:tcvu}. As $U$ approaches zero, 
the Gorkov method approaches the BCS mean-field result, differing by a factor of approximately $(4e)^{1/3}$, 
which is consistent with results obtained for Fermi gases. 
Notably, in the weak coupling limit, the Gorkov method, which incorporates the particle-hole channel, 
yields critical temperatures comparable to those from the NSR method, which also emphasizes the particle-particle channel. 
This similarity is worth further study.

Based on the results of Ref.~\cite{recent_fermionic,tma3,GorKov_Popov}, it is evident that their $G_0G$ method, 
which corresponds to our HGW method here, performs well in the unitarity region and aligns 
closely with both experimental and Monte Carlo results. This indicates the effectiveness of the 
negative $U$ Hubbard model in validating Fermi gas behavior. 
It also suggests that the GW method can provide accurate reference results in the small $U$ regime, 
where continuous Fermi gases are difficult to study via Monte Carlo simulations. 
Furthermore, the results show that HGW cannot use $W$
as a criterion for phase transition determination at small $U$. 
Using this criterion would lead to results that underestimate the true phase 
transition temperature, similar to the behavior observed in Fermi gas systems.

\subsection{\label{sec:gtau}Green's function at the Matsubara time axis }

We compare the Green's function obtained from different methods on the Matsubara time 
axis at the point 
$k=(\pi/2,\pi/2,\pi/2)$ for various doping levels, temperatures, and coupling strengths in an 
$8\times 8\times 8$ lattice. 
At high temperatures, the GW and HGW methods yield the normal phase solution. 
As the temperature decreases, the normal phase becomes unstable, leading to the superconducting phase characterized by long-range s-wave superconducting 
order and the breaking of the U(1) gauge symmetry. 
The comparison results between many-body methods and DQMC exhibit significant 
differences across different phases.

As shown in Fig.~\ref*{fig:gtau}, for $U=-4$ and away from half-filling, both the GW and HGW methods yield 
Green's functions that closely 
match the DQMC results in the normal phase.  
However, in the superconducting phase, both methods show deviations from the DQMC results. 
After applying the post-GW correction, which replaces the screened potential with physical responses, 
the imaginary-time Green's function shows significant improvement. 
 
Due to the presence of charge density waves (CDW) between the normal and superconducting phases at half-filling, 
we avoid regions where CDW may appear. In the normal phase, the HGW results consistently show significant deviations 
from the Monte Carlo data, while the GW method still provides reasonably accurate results. 
In the superconducting phase, the Green's functions from both 
GW and HGW agree with each other but deviate from the DQMC results, 
with the deviations decreasing as the temperature decreases. 
The post-GW method shows significant improvements near the 
critical region, and its results almost coincide with DQMC as the temperature continues to decrease. 
Near half-filling, the poor performance of HGW in the normal phase may be attributed to the asymmetry 
between the single-particle propagators for up and down spins in the particle-particle channel.

As the interaction strength increases, at $U=-8$ and away from half-filling GW and HGW exhibit distinct advantages in the normal phase. 
At high temperatures, the GW method delivers reliable results, while the HGW method demonstrates superior 
performance as the system approaches the superconducting transition temperature. 
Once the system enters the superconducting region, 
the post-GW method continues to perform well. 
Notably, we observe that, regardless of the parameters, 
the Green's functions obtained from GW and HGW in the superconducting phase are highly consistent. 
This may be attributed to the dominant role of the order parameter 
in the self-energy within the superconducting phase.

\subsection{\label{sec:NOSandprepair}Density of states and pseudogap }
    \begin{figure}[]
        \centering
        \includegraphics[width=1.\linewidth]{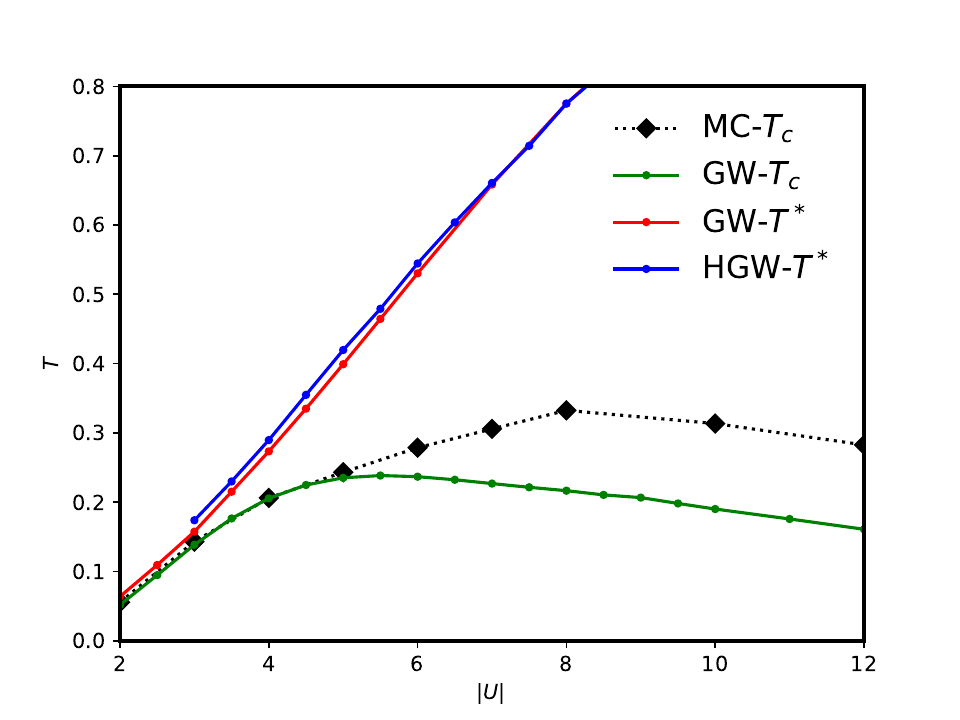}
        \caption{\label{fig:tstar} 
The temperature curve, determined by the presence of the superconducting phase obtained from the GW (red dotted line)  
and HGW (blue dotted line), is compared with the phase transition curve from DQMC (black diamonds line) and GW (green dotted line).}
        \end{figure}  
    \begin{figure*}[]
        \centering
        \includegraphics[width=1.\linewidth]{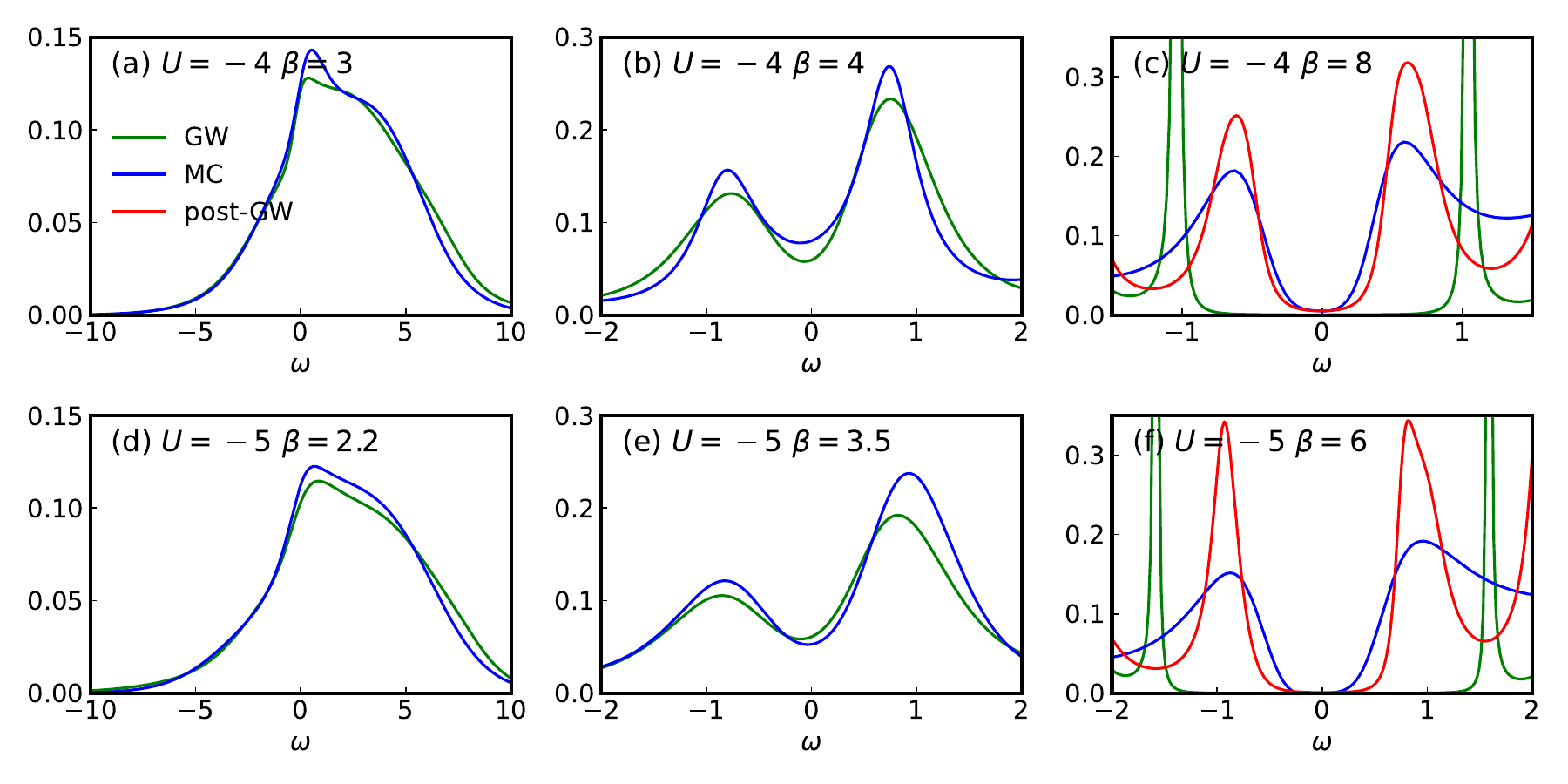}
        \caption{\label{fig:DOS} 
        The density of state in $8\times8\times8$ lattice. 
        These include DQMC results (blue line), GW results (green line) and post-GW results (red line). 
        All of the above plots correspond to a filling of $n = 0.50$. The first column corresponds to temperatures $T > T^*$, the second column to $T^* > T > T_c$, and the third column to $T_c > T$. 
        The specific analytical continuation method is described in 
        \cite{nevanlinna1,nevanlinna2,maxE1,maxE2}}
        \end{figure*}  
Starting from low temperatures and gradually increasing the temperature, we find that the GW and HGW methods continue to yield superconducting solutions (i.e., a non-zero order parameter) above the superconducting transition temperature $T_c$. These superconducting solutions persist until the temperature reaches a specific threshold, which we define as $T^*$.  
The $T^*$ obtained from the GW method is represented 
by the red dotted line in Fig.~\ref*{fig:tstar}. The $T^*$ obtained from the HGW method exhibits (the blue dotted line in Fig.~\ref*{fig:tstar}) is similar to that from the GW method, showing slightly higher values in the weak coupling regime and consistent results in the strong coupling regime. 
Notably, the $T^*$ obtained using this method increases 
monotonically with the magnitude of the coupling strength, in contrast 
to the results for the phase transition temperature $T_c$, which exhibits a maximum at a specific $U$ before gradually decreasing. As the interaction strength increases, the discrepancy between the two temperature curves gradually widens.

When local pairing fluctuations are strong, the GW and HGW methods may induce a broken-symmetry phase. 
In this phase, the single-particle density of 
states must exhibit a gap \cite{Tremblay_prepair}. However, both GW and HGW methods tend to overestimate the size of this gap (see Fig.~\ref*{fig:DOS}(c) and Fig.~\ref*{fig:DOS}(f)). Some theories propose the existence of two distinct excitation gaps \cite{review_Chen2,review_Chen1}. 
We define the temperature range where both the broken-symmetry and normal-phase solutions coexist as the metastable region. 
Given that the existence of the metastable region may be linked to electron pre-pairing \cite{capacityofpair,capacityofpair2}, 
the superconducting solutions obtained by GW and HGW above the superconducting transition temperature $T_c$ may be associated with pre-pairing. 
Considering these factors, we propose that $T^*$ is directly related to the temperature at which electron pre-pairing occurs.

To validate our hypothesis, we performed analytical continuation 
for both the GW method and the Monte Carlo method, analyzing the 
temperature curve of $T^*$ from the perspective of the energy gap. 
We employed the Nevanlinna method \cite{nevanlinna1,nevanlinna2} and the maximum entropy method \cite{maxE1,maxE2} to perform analytic continuation for the Green's function was obtained from the GW approximation and DQMC. 

Since the analysis is conducted on a discrete lattice, the Fermi surface may not coincide with the discrete momentum points.  
Therefore, we focused on analyzing the density of states:
\begin{equation}
    N(\omega) =\frac{1}{N}\sum_k A(\omega,k).
\end{equation}
At high temperatures, both the Monte Carlo and GW results exhibit a single peak in the density of states, 
with the GW results matching those of Monte Carlo (see Fig.~\ref*{fig:DOS}(a) and Fig.~\ref*{fig:DOS}(d)). In the regime where superconducting solutions 
begin to emerge but the normal phase has not yet undergone a full superconducting phase transition, 
the analytic continuation of DQMC and GW data reveals two distinct peaks, with the gap bottom remaining significantly above zero (see Fig.~\ref*{fig:DOS}(b) and Fig.~\ref*{fig:DOS}(e)). 
Upon fully entering the superconducting phase, GW predicts a gap larger than that obtained from DQMC, 
though both methods clearly show a superconducting gap. After applying the post-GW correction, the peak positions 
in the density of states align precisely with the DQMC results (see Fig.~\ref*{fig:DOS}(c) and Fig.~\ref*{fig:DOS}(f)). These results are consistent with the imaginary-time Green's function behavior was discussed earlier.

The above results demonstrate that when the temperature is above the $T^*$, 
both the normal-phase density of states from GW and DQMC do not exhibit a gap. Below the $T^*$, 
although a gap appears, it is not fully opened. It is only when the system enters the superconducting phase,  
below the superconducting transition temperature, that a true superconducting gap emerges.
The agreement of the post-GW improvements with Monte Carlo results after the correction suggests that the temperature 
at which the gap emerges in the GW solutions is directly related to preformed pairing. 
However, the limitations of the truncation in the GW method lead to an overestimation 
of the gap magnitude and the long-range correlations. 
Therefore, the broken phase results require further higher-order corrections to improve accuracy. 
Although the post-GW correction is a one-shot modification of the GW results 
and does not alter the temperature at which the gap first appears, 
it significantly improves the correction to the density of states. 
This also implies that the $T^*$ derived from GW or HGW retains physical significance, 
likely because the unphysical RPA treatment is excessively sensitive to strong local fluctuations induced by pre-pairing. 
Based on this, we conclude that the $T^*$ obtained from GW and HGW represents the temperature at which pre-pairing emerges in the system.
The increasing separation between the pairing temperature $T^*$ and 
the transition temperature $T_c$ indicates that the system transitions from the BCS region into the crossover region.

\subsection{\label{sec:green_and_ps} S-wave pair correlator}

\begin{figure*}[]
    \includegraphics[width=1\linewidth]{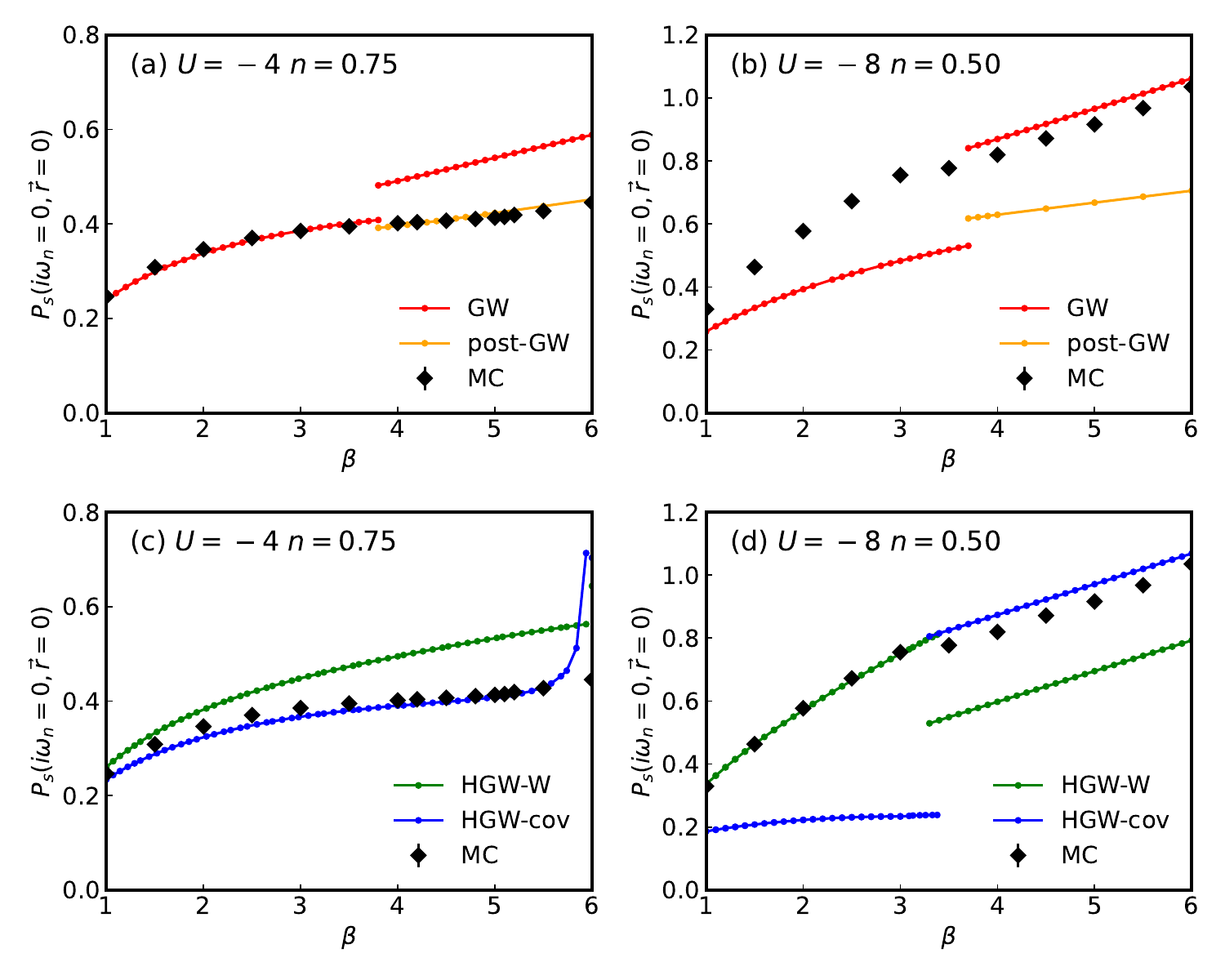}
    \caption{\label{fig:ps} 
    S-wave pair correlations in $\omega_n=0$ and $\vec r = 0$ with different methods in $8\times8\times8$ lattice. 
    The results include DQMC (black diamond line), 
    GW method from the normal phase (blue line), 
    HGW method from the normal phase by $W$ (orange line), 
    HGW method from the normal phase by covariance (green line), 
    GW method from the superconducting phase (purple dots line), 
    and post-GW from the superconducting phase (red dots line). }
    \end{figure*}

We also analyze the s-wave pair correlator from different methods at the coordinate space. 
Here, we fix the density and vary the temperature to compare the results. 
In the superconducting phase, the value of the pairing correlation function 
at the $\vec r =0$ is composed of two contributions. 
The first arises from the correlation function evaluated at nonzero momenta, 
and the second is a delta peak at zero momentum induced by the presence of the order parameter
\begin{equation}
    P_s^{\mathrm{SC}}(\vec r=0)=\frac{1}{N}\sum_{\vec k\neq 0}P_s^{\mathrm{SC}}(\vec k)+\langle \Delta^*\rangle\langle\Delta\rangle
\end{equation}
Combining these two contributions and performing a Fourier transform yields 
the value of the correlation function at the coordinate origin.

At $U = -4$ the covariant GW method provides two-particle correlation functions 
that agree well with DQMC results in the normal phase. 
However, in the superconducting phase, the GW method significantly overestimates 
the magnitude of the order parameter, leading to substantial deviations in the results. 
To address this, we introduced the post method to correct the Green's function, 
yielding final corrected results that closely match the Monte Carlo simulations.

For the intermediate coupling regime $U=-8$ the covariant GW results 
for the normal phase deviate significantly from the DQMC results. 
This indicates that correcting only the Green's function 
with the post-GW is insufficient, even though post-GW still performs well in calculating the 
single-particle Green's function, as shown in Fig.~\ref*{fig:ps}. 
In such cases, higher-order fully self-consistent methods are required, 
such as those incorporating the Gorkov contribution \cite{GorKov_Popov}. 
However, current computational constraints restrict the numerical implementation of these methods. 
Accurately calculating systems at strong coupling remains 
a significant challenge for advancing many-body physics.

For the HGW method, we use $W$ as the correlation function to compare with the correlation 
function obtained from the covariance calculation. At small coupling regime 
($U=-4$), such as in Fig.~\ref*{fig:ps}(b), the HGW normal phase solution remains stable 
as the temperature decreases but shows some deviation 
from DQMC results. In contrast, the covariance results align more closely with DQMC 
but diverge below the phase transition temperature predicted by DQMC. 
At the intermediate coupling regime ($U=-8$), the covariance results deviate significantly from the DQMC results. 
However, using $W$ as the pair correlation function yields results that agree well with the DQMC results. 
Only near the phase transition temperature are small discrepancies observed, 
because the coherence length exceeds the system size. A larger system will significantly 
reduce this discrepancy. This also explains why using $W$ as the criterion for determining the phase transition 
temperature yields reliable results in the intermediate-coupling regime. 
Additionally, it is worth noting that in the superconducting phase, 
the covariance results obtained from GW and HGW are nearly identical, 
similar to the previously calculated Green's functions.

The s-wave correlation function near the origin of coordinate space reflects short-range fluctuations. 
In the superconducting phase and weak-coupling regime, the post-GW method produces results 
that closely match the DQMC values after correcting the Green's function. 
This suggests that the superconducting phase calculations may be influenced by the unphysical RPA treatment 
of correlation functions, which fails to account for the suppression of the order parameter by true fluctuations. 
The screened potential $W$ in the RPA formalism leads to a significant overestimation of the order parameter and may result in incorrect identification of 
a superconducting phase, especially when stronger short-range correlations in the pairing gap are present.

\subsection{\label{sec:critical}Critical exponent}
\begin{table}
\caption{\label{tab:critical}%
    Comparison table of critical exponents from the 3D XY model and various methods.
We obtained the error by fitting the results for different fitting intervals, 
with the details provided in Appendix~\ref*{sec:detailofTc}. 
The D$\Gamma$A results are from Ref.~\cite{DGA1}.
    }
    \begin{ruledtabular}
        \begin{tabular}{l|p{1cm}|p{1cm}|p{2cm}|p{2cm}|p{1cm}}
            \textrm{ }&
            \textrm{mean field}&
            \textrm{$O(2)$ \newline (XY)}&
            \textrm{GW}&
            \textrm{HGW}&
            \textrm{ladder D$\Gamma$A }\\
            \colrule
            $\gamma$ & 1.00 & 1.33 & 1.31 $\pm$ 0.01 & 1.89 $\pm $ 0.02 & 1.90\\
            $\nu$ & 0.50 & 0.67 & 0.69 $\pm$ 0.02 & 0.92 $\pm $ 0.05 & 0.98\\
            \end{tabular}
    \end{ruledtabular}
\end{table}
The 3D s-wave superconductor state belongs to the same universality class as the 3D XY models \cite{CriticalExponents}.
The renormalization group theory indicates that the critical behavior is characterized by $\chi(T)=(T-T_c)^{-\gamma}$ and $\xi = (T-T_c)^{-\nu}$. 
In mean field theory, the critical exponents are given as $\gamma = 1$ and $\nu = \frac{1}{2}$, 
which are also consistent with results obtained from DMFT. 
However, the Monte Carlo calculations for the 3D XY model yield critical exponents of 
$\gamma \approx 1.32$ and $\nu \approx 0.67$ \cite{3DXY,3DXY_book}. 
Therefore, it is essential to consider contributions beyond mean field theory to account for nonlocal spatial correlations, 
which can be addressed through the covariance GW framework.

Starting from the normal phase, we gradually reduce the temperature and, based on this, calculate the two critical exponents mentioned above.
According to the 3D Landau theory, in the normal phase, 
we determine the correlation function for fluctuations in real space as follows:
\begin{equation}
    P_s(\vec r) \sim \frac{e^{-|\vec r|/\xi(T)}}{|\vec r|} \ \ \vec r>>0.\label{critical length}
\end{equation}
The correlation length is obtained by fitting the pair correlation function derived in real space 
according to Eq.~(\ref*{critical length}). The details are presented in Appendix~\ref*{sec:detailofTc}.
Due to the finite size of the system, we consider the critical region to correspond to the range 
where the correlation length is comparable in magnitude to the system size, 
truncating at the point where the correlation length significantly exceeds the system size. 

Due to numerical precision issues, we select multiple sets of fitting 
intervals for the real-space correlation length and critical region, 
obtaining several values for the critical exponents. 
Despite changes in the fitting procedures and critical exponents, our results show the robustness of the estimated critical temperatures, consistent with the results from previous methods, which validates the reliability of these two calculations. 

The critical exponent's results are summarized in Table.~\ref*{tab:critical}. 
The GW results are obtained for the parameters $U = -4, n = 0.5$, 
and the HGW results are obtained for $U = -8, n = 0.5$. 
Notably, the GW method consistently yields critical exponents 
similar to those of the 3D XY model, which $\gamma=1.31 \pm 0.01$ and $\nu= 0.69 \pm 0.02$. However, the $W$ obtained from the HGW method significantly overestimates the critical exponents as $\gamma=1.89 \pm 0.02$ and $\nu= 0.92 \pm 0.05$.  
The discrepancy in the critical exponent results for the HGW method may 
arise from the asymmetry of Green's function in the polarization bubble, as discussed earlier.

\section{\label{sec:conclusion}Conclusion and Discussion}

In summary, we calculated different non-perturbative manybody theories in the three-dimensional Hubbard model. Specifically, 
we employ the GW and HGW methods to compute transition temperature and critical exponents 
starting from the normal phase. In the superconducting phase, the post-GW method is utilized to 
calculate Green's function and density of states. Additionally, we compute Green's function 
and pairing correlation functions using these approaches. Our results are compared with DQMC simulations, 
offering a reference for many-body theories without the delicacies of 
discretization and renormalization in Fermi gas systems.

In the normal phase, we find that the results from the self-consistent GW method are close to 
DQMC results in the weak to intermediate coupling regime 
including Green's functions, pairing correlation functions, 
and transition temperatures. It also yields critical exponents 
close to those of the 3D XY model. 
However, deviations arise as the interaction strength increases. 
In contrast, the HGW method provides transition temperatures that are 
in good agreement with DQMC, particularly at the intermediate coupling regime 
and away from half-filling.  Both methods capture the BCS-BEC crossover behavior, 
where the transition temperature initially increases with 
interaction strength, reaching a maximum, and then decreasing as the 
interaction strength further increases, but both exhibit quantitative deviations 
in the crossover region approaching the BEC limit.
The different applicability ranges of these methods result from different approximations: 
the GW method is based on vertex truncation, while the HGW method 
truncates high-order connected correlators. 

Starting from low temperatures, we obtain a superconducting solution characterized by a 
non-zero anomalous Green's function. We find that both the standard GW and HGW methods exhibit 
deviations from DQMC results when calculating Green's function in this phase. 
The results indicate that the emergence of the anomalous Green's function makes the RPA insufficient for self-consistently 
calculating two-particle correlation functions, leading to the violation of the FDT. 
Consequently, the GW truncation disrupts the relationship between the RPA-like screened potential 
$W$ and the pairing correlations. 
To improve this, we employ the covariance method in the superconducting phase, 
which is based on the physical response to external sources, preserving both the FDT and WTI. 
Furthermore, we replace the screened potential 
$W$ in the GW approximation with the covariant correlation function. 
This approach yields the post-GW method, which significantly enhances the accuracy 
of Green's function, density of states, and correlation functions, especially near the critical regime.

Additionally, above the superconducting transition temperature, both the GW and HGW methods still yield superconducting solutions up to a temperature $T^*$. 
We associate this temperature curve with the pairing temperature, which suggests the appearance of the pseudogap. 
Both the HGW and GW methods yield similar curves, 
with the temperature increasing monotonically with interaction strength. 
By analyzing the density of states, we observe that no gap is present above the $T^*$ 
but emerges below it. 
The superconducting gap, however, only appears when the temperature is below 
the superconducting transition temperature. 
In the pseudogap region between $T^*$ and $T_c$, localized paired fermions exist, but coherence between Cooper pairs is absent.
This explains why the GW and HGW methods can provide superconducting solutions in this region, but the system does not enter a true superconducting state. 

In this work, both the GW and HGW methods are confined to the particle-particle channel. Notably, 
in the weak coupling limit, the Gorkov method, which includes the particle-hole channel, yields critical 
temperatures similar to those from the GW method. 
This phenomenon currently lacks a clear explanation and requires further investigation. 
In the crossover region near the BEC limit,  
the phase diagrams from both the GW and HGW methods show significant deviations from DQMC results. 
To address this, it may be necessary to go beyond the GW method by 
considering higher-order Feynman diagrams \cite{GorKov_Popov,recent_fermionic}.

\begin{acknowledgments}
We thank Ruitao Xiao, Zhipeng Sun, Huaqing Huang, Ziyu Li for helpful discussions and technical assistance.
This work is supported by the National Natural Science Foundation of China (Grants No. 12074006 and 91736208) 
and the High-performance Computing Platform of Peking University.
\end{acknowledgments}

\appendix

\section{NAMBU SPACE AND DYSON-SCHWINGER EQUATION}
Following the notation of Nambu \cite{Nambu}, it will be convenient to introduce a 
two-component notation for the electron field
\begin{equation}
    \Psi(1) = \left(\begin{array}{l}
        \psi_{\uparrow}(1)\\
        \psi^{*}_{\downarrow}(1)
    \end{array}\right) \ 
    \mathrm{and} \
    \Psi^*(1) = \left(\begin{array}{ll}
        \psi^*_{\uparrow}(1) &
        \psi_{\downarrow}(1)
    \end{array}\right).
\end{equation}
Thus the one-body Green's function defined by ensemble average can be 
written as 
\begin{equation}
\begin{aligned}
    \underline{G}(1,2)
    =&\langle \Psi^{*}(2)\Psi(1)\rangle\\
    =&\left(\begin{array}{ll}
        \langle\psi^*_\uparrow(2)\psi_\uparrow(1)\rangle & \langle\psi_\downarrow(2)\psi_\uparrow(1)\rangle\\
        \langle\psi^*_\uparrow(2)\psi^*_\downarrow(1)\rangle & \langle\psi_\downarrow(2)\psi^*_\downarrow(1)\rangle
    \end{array}\right).
\end{aligned}
\end{equation}
Further, according to the invariance of the functional integral measure
$\mathcal{D}[\psi,\psi^{*}]$ under the infinitesimal variation of field $\psi$, $\psi^*$, 
one can obtain the equality
\begin{equation}
    \int \mathcal{D}[\psi,\psi^{*}]\frac{\delta }{\delta \Psi^{*}(2)}\left(\Psi^{*}(1)
    e^{-S[\psi,\psi^{*};J,J^{*}]}\right)=0.\label{EOM}
\end{equation}
Substituting the action Eq.~(\ref*{general action}) into the equality, one obtaines the Dyson-Schwinger 
equation of motion:
\begin{widetext}
\begin{equation}
\begin{aligned}
    \delta(1,2)=&\int d(3)T(1,3)\langle \psi^{*}_{\uparrow}(2)\psi_{\uparrow}(3)\rangle
    -\int d(3)V(1,3)\langle \psi^{*}_{\uparrow}(2)\psi^{*}_{\downarrow}(1)
    \psi_{\downarrow}(3)\psi_{\uparrow}(3)\rangle 
    + J^{+}(1)\langle\psi^{*}_{\uparrow}(2)\psi^{*}_{\downarrow}(1)\rangle\\    
    0=&\int d(3)T(1,3)\langle \psi_{\downarrow}(2)\psi_{\uparrow}(3)\rangle
    -\int d(3)V(1,3)\langle \psi_{\downarrow}(2)\psi^{*}_{\downarrow}(1)
    \psi_{\downarrow}(3)\psi_{\uparrow}(3)\rangle 
    + J^{+}(1)\langle\psi_{\downarrow}(2)\psi^{*}_{\downarrow}(1)\rangle\\  
    0=&-\int d(3)T(3,1)\langle \psi^{*}_{\uparrow}(2)\psi^{*}_{\uparrow}(3)\rangle
    -\int d(3)V(1,3)\langle \psi^{*}_{\uparrow}(2)\psi_{\uparrow}(1)
    \psi^{*}_{\uparrow}(3)\psi^{*}_{\downarrow}(3)\rangle 
    + J^{-}(1)\langle\psi^{*}_{\uparrow}(2)\psi_{\uparrow}(1)\rangle\\    
    \delta(1,2)=&-\int d(3)T(3,1)\langle \psi_{\downarrow}(2)\psi^{*}_{\uparrow}(3)\rangle
    -\int d(3)V(1,3)\langle \psi_{\downarrow}(2)\psi_{\uparrow}(1)
    \psi^{*}_{\uparrow}(3)\psi^{*}_{\downarrow}(3)\rangle 
    + J^{-}(1)\langle\psi_{\downarrow}(2)\psi_{\uparrow}(1)\rangle.\\ 
\end{aligned}\label{Dyson motion equation}
\end{equation}
\end{widetext}
Through the definition of Green's functional Eq.~(\ref*{Nambu Green's function}), one can 
obtains the derivative of $G$ with respect to $J$ and $J^{*}$:
\begin{equation}
\begin{aligned}
    \frac{\delta \langle \psi^{B}(2)\psi^{A}(1)\rangle}{\delta J^{+}(3)}
    =&\langle \psi^{B}(2)\psi^{A}(1)\psi^{*}_{\uparrow}(3)\psi^{*}_{\downarrow}(3)\rangle\\
    &-\langle \psi^{B}(2)\psi^{A}(1)\rangle
    \langle \psi^{*}_{\uparrow}(3)\psi^{*}_{\downarrow}(3)\rangle\\
    \frac{\delta \langle \psi^{B}(2)\psi^{A}(1)\rangle}{\delta J^{-}(3)}
    =&\langle \psi^{B}(2)\psi^{A}(1)\psi_{\downarrow}(3)\psi_{\uparrow}(3)\rangle\\
    &-\langle \psi^{B}(2)\psi^{A}(1)\rangle
    \langle \psi_{\downarrow}(3)\psi_{\uparrow}(3)\rangle,
\end{aligned}\label{definition of Green}
\end{equation}
where the $A$, $B$ label the spin and charge. 
Combining Eq.~(\ref*{definition of Green}) and Dyson-Schwinger equation of motion 
Eq.(\ref*{Dyson motion equation}), one can obtain Eq.~(\ref*{Dyson-Schwinger}) in matrix form.
\section{\label{sec:detailsofGW}DETAILS OF DERIVING HGW EQUATIONS AND GW EQUATIONS}

\subsection{Derivation of GW equations}
Note that the functional derivative $\frac{\delta G}{\delta J^a}$ can be rewritten as 
\begin{equation}
    \frac{\delta \underline{G}(1,2)}{\delta J^{a}(3)}
    =-\int \mathrm{d}(45)\sum_b\underline{G}(1,4)\underline{\Lambda}^b(4,5,6)
    \underline{G}(5,2)\frac{\delta v^{b}(6)}{\delta J^{a}(3)},
\end{equation}
where $\underline{\Lambda}^a(1,2,3)$ represents Hedin's vertex function, defined as
\begin{equation}
    \underline{\Lambda}^a(1,2,3)\equiv\frac{\delta \underline{G}^{-1}(1,2)}{\delta v^{a}(3)}.
\end{equation}
Using the form of the single-particle effective potential  
from Eq.~(\ref*{v format}), we can derive
\begin{widetext}
\begin{equation}
\begin{aligned}
    \frac{\delta v^{a}(1)}{\delta J^{b}(2)}
    =\delta^{ab}\delta(1,2)
    +\int d(456)\sum_{c}V(1,3)\mathrm{Tr}\left[
        \underline{\sigma}^{\bar a}
        \underline{G}(3,4)
    \frac{\delta \underline{G}^{-1}(4,5)}{\delta v^{c}(6)}
    \underline{G}(5,3)
    \right]\frac{\delta v^{c}(6)}{\delta J^{b}(2)}.
\end{aligned}\label{vequation}
\end{equation}
\end{widetext}
The screened dynamical potential $W_{G}$ is defined as 
\begin{equation}
    W_{\mathrm{G}}^{ab}(1,2) = \int \mathrm{d}(3) \frac{\delta v^a(1)}{\delta J^b(3)}V(2,3).
\end{equation}
According to Eq.~(\ref*{vequation}), we can obtain the equation for 
$W$
\begin{equation}
\begin{aligned}
        W^{ab}_G(1,2)=&V(1,2)\delta^{ab}\\ &+\int \mathrm{d}(34)\sum_c V(1,3)\Pi^{ac}(3,4)W^{cb}(4,2).
\end{aligned}
\end{equation}
The corresponding polarization function is given by:
\begin{equation}
    \Pi^{ab}(1,2)=\int d(34)\mathrm{Tr}\left[
    \underline{\sigma}^{\bar a}
    \underline{G}(1,3)
    \underline{\Lambda}^{b}(3,4,2)
    \underline{G}(4,1)\right].
\end{equation}
Combining the above equations, we obtain Hedin's equations
\begin{subequations}
    \begin{align}
        \underline{G}^{-1}(1,2)=&\underline{H}^{-1}(1,2)-\underline{\Sigma}(1,2),\\
        \underline{\Sigma}(1,2)=&-\sum_{ab}\underline{\sigma}^{a}
        \underline{G}(1,4)\underline{\Lambda}^{b}(4,2,5)W^{b\bar a}(5,3),\\
        W^{ab-1}(1,2)=&V^{-1}(1,2)I^{ab}-\Pi^{ab}(1,2),\\
        \Pi^{ab}(1,2)=&\int d(34)\mathrm{Tr}\left[
    \underline{\sigma}^{\bar a}
    \underline{G}(1,3)
    \underline{\Lambda}^{b}(3,4,2)
    \underline{G}(4,1)\right].
    \end{align}\label{eq:Hedin}
\end{subequations}
The above equations are entirely rigorous and accurate since no approximations have been made. However, they cannot be solved numerically without appropriate truncations. 
By applying the truncation scheme in Eq.~(\ref*{GGW truncation}) we can derive the GW equations.

\subsection{Derivation of HGW equations}
According to the truncation method described in Sec.~\ref*{sec:HGWandGGW}, we can obtain
\begin{equation}
    \frac{\delta \underline{G}(1,2)}{\delta J^{a}(3)} = -\int\mathrm{d}(45)\underline{H}(1,4)
    \frac{\delta \underline{H}^{-1}(4,5)}{\delta J^{a}(3)}\underline{G}(5,2).
    \label{HGW truncation}
\end{equation}
Starting from the Dyson-Schwinger equation of motion Eq.~(\ref*{Dyson-Schwinger})
and applying the truncation scheme Eq.~(\ref*{HGW truncation}), one can derive the HGW equations. 
The process begins by taking derivatives of Eq.~(\ref*{H format}) with respect to the external sources 
$J^{+}$ and $J^{-}$, yielding the following expression
\begin{equation}
    \frac{\delta \underline{H}^{-1}(1,2)}{\delta J^{a}(3)}
    =\delta(1,2)\sum_{b}\underline{\sigma}^{b}\frac{\delta v^{b}(1)}{\delta J^{a}(3)}.\label{Hderivative}
\end{equation}

Using the form of the single-particle effective potential  
from Eq.~(\ref*{v format}), we can derive
\begin{equation}
\begin{aligned}
    \frac{\delta v^{a}(1)}{\delta J^{b}(2)}
    =&\delta(1,2)\delta^{ab}-\int d(3)V(1,3)\mathrm{Tr}\left[\underline{\sigma}^{\bar a}
        \frac{\delta \underline{G}(3,3)}{\delta J^{b}(2)}
    \right].
\end{aligned}\label{vderivative}
\end{equation}
Substituting Eq.~(\ref*{HGW truncation}) into Eq.~(\ref*{vderivative}), we obtain
\begin{equation}
\begin{aligned}
    \frac{\delta v^{a}(1)}{\delta J^{b}(2)}=&
    \delta(1,2)\delta^{ab}\\
    &+\int d(34) \sum_{c}V(1,3)\mathrm{Tr}\left[
        \Pi_{\mathrm{H}}^{\bar ac}(3,4)\frac{\delta v^{c}(4)}{\delta J^{b}(2)}
    \right],
\end{aligned}
\end{equation}
where the polarization function is given by
\begin{equation}
    \Pi^{ab}_{\mathrm{H}}(1,2)=\mathrm{Tr}\left[\underline{\sigma}^{\bar a}\underline{H}(1,2)
    \underline{\sigma}^{b}\underline{G}(2,1)\right].
\end{equation}

Next, by substituting both Eqs.~(\ref*{HGW truncation}) and (\ref*{Hderivative}) 
into the Dyson-Schwinger equation (Eq.~(\ref*{Dyson-Schwinger})), we obtain the following expression
\begin{equation}
\begin{aligned}
    \underline{I}\delta(1,2)=&\int\mathrm{d}(3) \underline{H}^{-1}(1,3)\underline{G}(3,2)\\
    &+\int\mathrm{d}(3)\sum_{a}V(1,3)\underline{\sigma}^{+}\underline{H}(1,4)
    \underline{\sigma}^{a}\underline{G}(4,2)\frac{\delta v^a(4)}{\delta J^{-}(2)}\\
    &+\int\mathrm{d}(3)\sum_{a}V(1,3)\underline{\sigma}^{-}\underline{H}(1,4)
    \underline{\sigma}^{a}\underline{G}(4,2)\frac{\delta v^a(4)}{\delta J^{+}(2)}.
\end{aligned}
\end{equation}
This equation can be rewritten as 
\begin{equation}
    \underline{G}^{-1}(1,2)=\underline{H}^{-1}(1,2)-\underline{\Sigma}_{\mathrm{H}}(1,2),
\end{equation}
where the self-energy function $\underline{\Sigma}_{\mathrm{H}}$ is given by 
\begin{equation}
    \underline{\Sigma}_{\mathrm{H}}(1,2)=-\sum_{a,b=\pm}\underline{\sigma}^{a}
    \underline{H}(1,4)\underline{\sigma}^{b}(4,2,5)W^{b\bar a}_{\mathrm{H}}(5,3).
\end{equation}
Additionally, the screened dynamical potential $W_{\mathrm{H}}$ is defined as 
\begin{equation}
    W_{\mathrm{H}}^{ab}(1,2) = \int \mathrm{d}(3) \frac{\delta v^a(1)}{\delta J^b(3)}V(2,3).\label{HGWW}
\end{equation}
Combining Eqs.~(\ref*{vderivative}) and (\ref*{HGWW}), , we arrive at the following equation for $W_{H}$
\begin{equation}
    \underline{W}_{\mathrm{H}}^{ab} = V(1,2)I^{ab} + \int \mathrm{d}(34)\sum_{c}V(1,3)\Pi_{\mathrm{H}}^{ac}(3,4)W^{cb}_{\mathrm{H}}(4,2).
\end{equation}
This can be rewritten as
\begin{equation}
    W^{ab-1}_{\mathrm{H}}(1,2)=V^{-1}(1,2)I^{ab}-\Pi^{ab}_{\mathrm{H}}(1,2).
\end{equation}
By combining all of the above equations, we obtain the full set of the HGW equations.

\section{\label{apd:covariance}DETAILS OF DERIVING COVARIANT EQUATIONS}
Based on the expression for the dynamical screening potential, 
after introducing a external source, 
its derivative with respect to the source can be written as
\begin{equation}
\begin{aligned}
    &\Lambda^{ab}_{\phi,\mathrm{W}}(1,2,3)=\frac{\delta W^{ab}(1,2)}{\delta \phi(3)}\\
    &=\int d(45)\sum_{cd}W^{ac}(1,4)\frac{\delta W^{cd-1}(4,5)}{\delta \phi(3)}
    W^{db}(5,2)\\
    &=\int d(45)\sum_{cd}W^{ac}(1,4)\Gamma^{cd}_{\phi,\mathrm{W}}(4,5,3)
    W^{db}(5,2).
\end{aligned}
\end{equation}
Furthermore, based on the inverse expression, it can be written as
\begin{equation}
\begin{aligned}
    &\Gamma^{ab}_{\phi,\mathrm{W}}(1,2,3)
    =-\frac{\delta \Pi^{ab}(1,2)}{\delta \phi(3)}\\
    &=-\mathrm{Tr}\left[\underline{\sigma}^{\bar a}\underline{\Lambda}_{\phi}(1,2)
    \underline{\sigma}^{b}\underline{G}(2,1)
    +\underline{\sigma}^{\bar a}\underline{G}(1,2)
    \underline{\sigma}^{b}\underline{\Lambda}_{\phi}(2,1)\right].
\end{aligned}
\end{equation}
Combining with Eq.~(\ref*{Covariant vertex equation}), 
this is the complete expression for the covariant scheme. 

For HGW, according to HGW equations Eq.~(\ref*{HGW equation})
\begin{equation}
    \begin{aligned}
    \underline{\Lambda}^{H}_{\phi}(1,2,3)
    =&-\int\mathrm{d}(456)\underline{G}(1,4)
        (\gamma+\underline{\Gamma}_{\mathrm{H}})(4,5,3)\underline{G}(5,2),
    \end{aligned}
\end{equation}
and the vertices of covariance:
\begin{subequations}
    \begin{align}
        &\underline{\Gamma}_{\phi,\mathrm{MT}}(1,2,3) = \sum_{a,b}\underline{\sigma}^{a}
        \underline{\Lambda}^{\mathrm{H}}_{\phi}(1,2,3)\underline{\sigma}^{b}W^{b\bar a}(2,1),\\
        &\underline{\Gamma}_{\phi,\mathrm{AL}}(1,2,3) = \sum_{a,b}\underline{\sigma}^{a}
        \underline{H}(1,2)\underline{\sigma}^{b}\Lambda_{\phi,W,\mathrm{H}}^{b\bar a}(2,1,3),
    \end{align}
\end{subequations}
where $\Gamma^{ab}_{\phi,W,\mathrm{H}}$ has the form:
\begin{equation}
\begin{aligned}
    &\Gamma^{ab}_{\phi,W,\mathrm{H}}
    =\\
    &-\mathrm{Tr}\left[\underline{\sigma}^{\bar a}\underline{\Lambda}^{H}_{\phi}(1,2)
    \underline{\sigma}^{b}\underline{G}(2,1)
    +\underline{\sigma}^{\bar a}\underline{H}(1,2)
    \underline{\sigma}^{b}\underline{\Lambda}_{\phi}(2,1)\right].
\end{aligned}
\end{equation}

\section{IMPLEMENT IN THE THREE-DIMENSIONAL HUBBARD MODEL}

\subsection{Discretization of Matsubara imaginary time}
The discretized Matsubara time Matsubara action has the form
\begin{equation}
\begin{aligned}
    S_{M}[\psi,\psi^*]=&\sum_{l=0}^{M-1}\sum_{\sigma=\uparrow,\downarrow}\sum_{i}
    \psi^*_{i\sigma}(\tau_l)\left[\psi_{i\sigma}(\tau_{l+1})-\psi_{i\sigma}(\tau_{l})\right]\\
    &+\Delta\tau\sum_{l=0}^{M-1}\mathcal{H}[\psi^*_{i\sigma}(\tau_l),\psi_{i\sigma}(\tau_{l})].\label{Matsubara action}
\end{aligned}
\end{equation}
Here $M$ is the number of Matsubara time slices, and $\Delta \tau \equiv \beta / M$ is the time step. 
The $l$ labels the discretized Matsubara time, so $\tau_l\equiv l\Delta\tau$. The function $\mathcal{H}$ is 
obtained by substituting $\psi^*_{i\sigma}(\tau_l)$, $\psi_{i\sigma}(\tau_{l})$ for 
$c^\dagger_{i\sigma}$, $c_{i\sigma}$ in Hamiltonian $\hat{\mathcal{H}}$, Eq.~(\ref*{Cooper Hamiltonian}). 

Comparing the Matsubara action Eq.~(\ref*{Matsubara action}) with 
the form Eq.~(\ref*{generalized action}), one can obtain the expression for the kinetic term $T$
\begin{equation}
\begin{aligned}
    T(1,2)=&-\frac{1}{\Delta\tau}\delta_{i_1i_2}\left(\delta_{l_1,l_2-1}-\delta_{l_1,l_2}\right)\\
    &-t_{i_1i_2}\delta_{l_1,l_2}+\mu\delta_{i_1i_2}\delta_{l_1,l_2},
\end{aligned}\label{T space term}
\end{equation}
and the interaction potential $V$
\begin{equation}
    V(1,2)=-\delta_{i_1i_2}\delta_{l_1,l_2}|U|.
\end{equation}
Where the labels $1$, $2$ denote $\left(i_1,\tau_{1}\right)$, $\left(i_2,\tau_{2}\right)$, respectively. 
The hopping strength $t_{i_1i_2}$ only considers the nearest neighbor term.

\subsection{Fourier transformation for Hubbard model}
For a lattice with translation symmetries, using the discrete Fourier transformation to solve the equations in momentum space will simplify the computation. 
For a fermionic array $X_{\mathrm{F}}$ which is antiperiodic in Matsubara time takes the form
\begin{equation}
    X_{\mathrm{F}}(1,2)=\frac{1}{\mathcal{N}}\sum_{k}X_{\mathrm{F}}(k)\epsilon_{\mathrm{F}}(k,1-2),
\end{equation}
and the Bosonic array $X_{\mathrm{B}}$ takes the form
\begin{equation}
    X_{\mathrm{B}}(1,2)=\frac{1}{\mathcal{N}}\sum_{k}X_{F}(k)\epsilon_{\mathrm{B}}(k,1-2).
\end{equation}
Here $\mathcal{N}=\beta N$ and $k=(n,\vec k)$, $N$ is the number of lattice sites and 
$n$ takes the integer value from $0$ to $M-1$. The transformation kernels $\epsilon_{\mathrm{F}}$ 
and $\epsilon_{\mathrm{B}}$ can be written as 
\begin{equation}
    \epsilon_{\mathrm{F}}(k,1-2) \equiv e^{-\mathrm{i}\pi\frac{2n+1}{M}(l_1-l_2)}e^{i\vec k \cdot(\vec x_1-\vec x_2)}
    \label{fermion transformation}
\end{equation}
\begin{equation}
    \epsilon_{\mathrm{B}}(k,1-2) \equiv e^{-\mathrm{i}\pi\frac{2n}{M}(l_1-l_2)}e^{i\vec k \cdot(\vec x_1-\vec x_2)}
    \label{bosonic transformation}
\end{equation}
For the kinetic term $T$, it's transformation can be written as 
\begin{subequations}
    \begin{align}
        \mathfrak{F}(T(1,2))(k)
        =&-\frac{1}{\Delta \tau}\left(e^{-\mathrm{i}\pi(2n+1)/M}-1\right)-\varepsilon(\vec k)+\mu,\\
        \mathfrak{F}(T(2,1))(k)
        =&-\frac{1}{\Delta \tau}\left(e^{\mathrm{i}\pi(2n+1)/M}-1\right)-\varepsilon(-\vec k)+\mu.\\        
    \end{align}
\end{subequations}
Here, for the three-dimensional Hubbard model,
 $\varepsilon(\vec k)=-2t\left(\cos k_x+\cos k_y+\cos k_z\right)$ is 
 the noninteracting dispersion with t the nearest-neighbor hopping strength. The interaction 
 potential $V$ in momentum space can be written as $V(k)=U$.

\subsection{HGW and GW equations for thr Hubbard model}
In HGW equations, the $\underline{H}$, $\underline{G}$ and $\underline{\Sigma}$
are the fermionic arrays, and $W$, $\Pi$ are bosonic arrays. In a translation invariance lattice 
with periodic boundary condition, substituting Eqs.~(\ref*{fermion transformation}) and 
(\ref*{bosonic transformation}) into the $HGW$ equations Eqs.~(\ref*{HGW equation}), one can 
obtain
\begin{subequations}
    \begin{align}
        \underline{G}^{-1}(k)=&\underline{H}^{-1}(k)-\underline{\Sigma}_{\mathrm{H}}(k),\\
        \underline{\Sigma}_{\mathrm{H}}(k)=&-\frac{1}{\mathcal{N}}\sum_{a,b=\pm}\sum_{q}\underline{\sigma}^{a}
        \underline{H}(k+q)\underline{\sigma}^{b}W^{b\bar a}_{\mathrm{H}}(q),\\
        W^{ab-1}_{\mathrm{H}}(k)=&V^{-1}(k)I^{ab}-\Pi^{ab}_{\mathrm{H}}(k),\\
        \Pi^{ab}_{\mathrm{H}}(k)=&\frac{1}{\mathcal{N}}\sum_{q}\mathrm{Tr}\left[\underline{\sigma}^{\bar a}\underline{H}(k+q)
        \underline{\sigma}^{b}\underline{G}(q)\right],
    \end{align}
\end{subequations}
with 
\begin{equation}
    \underline{H}^{-1}(k)=\underline{T}(k)-\frac{U}{\mathcal{N}}
    \sum_{a}\sum_{q}\underline{\sigma}^{a}\mathrm{Tr}\left[\underline{\sigma}^{\bar a}
    \underline{G}(q)\right].
\end{equation}

According to the same process, the $GGW$ equations in momentum space can be written as 
\begin{subequations}
    \begin{align}
        \underline{G}^{-1}(k)=&\underline{H}^{-1}(k)-\underline{\Sigma}_{\mathrm{G}}(k),\\
        \underline{\Sigma}_{\mathrm{G}}(k)=&-\frac{1}{\mathcal{N}}\sum_{a,b=\pm}\sum_{q}\underline{\sigma}^{a}
        \underline{G}(k+q)\underline{\sigma}^{b}W^{b\bar a}_{\mathrm{G}}(q),\\
        W^{ab-1}_{\mathrm{G}}(k)=&V^{-1}(k)I^{ab}-\Pi^{ab}_{\mathrm{G}}(k),\\
        \Pi^{ab}_{\mathrm{G}}(k)=&\frac{1}{\mathcal{N}}\sum_{q}\mathrm{Tr}\left[\underline{\sigma}^{\bar a}\underline{G}(k+q)
        \underline{\sigma}^{b}\underline{G}(q)\right].
    \end{align}
\end{subequations}
\subsection{Covariant equations for the Hubbard model}
Since the covariant equations involve three-points correlation and vertex functions, 
we first make ansatz 
\begin{equation}
    X(1,2,3)=\frac{1}{\mathcal{N}^2}\sum_{p,q}X(p,q)\epsilon_{F}(p,1-2)\epsilon_{B}(q,1-3).
\end{equation}
Similarly to the above derivation of the GW and HGW equations one gets
\begin{equation}
    \underline{\Lambda}_{\phi}(p,q)=-\underline{G}(p+q)\underline{\Gamma}_{\phi}(p,q)
    \underline{G}(p),
\end{equation}
\begin{equation}
\begin{aligned}
\underline{\Gamma}_\phi(p,q)=&\underline{\gamma}_\phi(p,q)+\underline{\Gamma}_{\phi,\mathrm{H}}(p,q)\\
    &+\underline{\Gamma}_{\phi,\mathrm{MT}}(p,q)+\underline{\Gamma}_{\phi,\mathrm{AL}}(p,q),
\end{aligned}
\end{equation}
\begin{equation}
    \underline{\Gamma}_{\phi,\mathrm{H}}(p,q) 
    =-\frac{U}{\mathcal{N}}\sum_{a}\sum_{k}\underline{\sigma}^{a}\mathrm{Tr}
    \left[\underline{\sigma}^{\bar a}
    \underline{\Lambda}_{\phi}(k,q)\right],
\end{equation}
\begin{equation}
    \underline{\Gamma}_{\phi,\mathrm{MT}}(p,q) = \frac{1}{\mathcal{N}}\sum_{a,b}\sum_{k}\underline{\sigma}^{a}
    \underline{\Lambda}_{\phi}(p+k,q)\underline{\sigma}^{b}W^{b\bar a}(k),
\end{equation}
\begin{equation}
    \underline{\Gamma}_{\phi,\mathrm{AL}}(p,q) = \frac{1}{\mathcal{N}}\sum_{a,b}\sum_{k}\underline{\sigma}^{a}
    \underline{G}(p+k+q)\underline{\sigma}^{b}\Lambda_{\phi,\mathrm{W}}^{b\bar a}(k,q).
\end{equation}

For a given set of parameters $U$, $\mu$, $\beta$ and lattice size parameters $N$, $M$, one can 
solve the GW Eqs.~(\ref*{GGW equation}) and HGW Eqs.~(\ref*{HGW equation}) self-consistently to 
obtain Green's functions. With Green's function and bare vertices, one can  
solve linear Eq.~(\ref*{Covariant chi}) to get two-body correlation functions. 
Furthermore, for GW, one can use the two-body correlation functions obtained from the covariance method 
and substitute them into Eq.~(\ref*{post equation}) to do a one-shot calculation, then one obtains 
so-call post-GW solution.

\section{\label{sec:detailofTc}Detail of finite size scaling and critical exponent}
\begin{figure*}[]
    \includegraphics[width=1.0\linewidth]{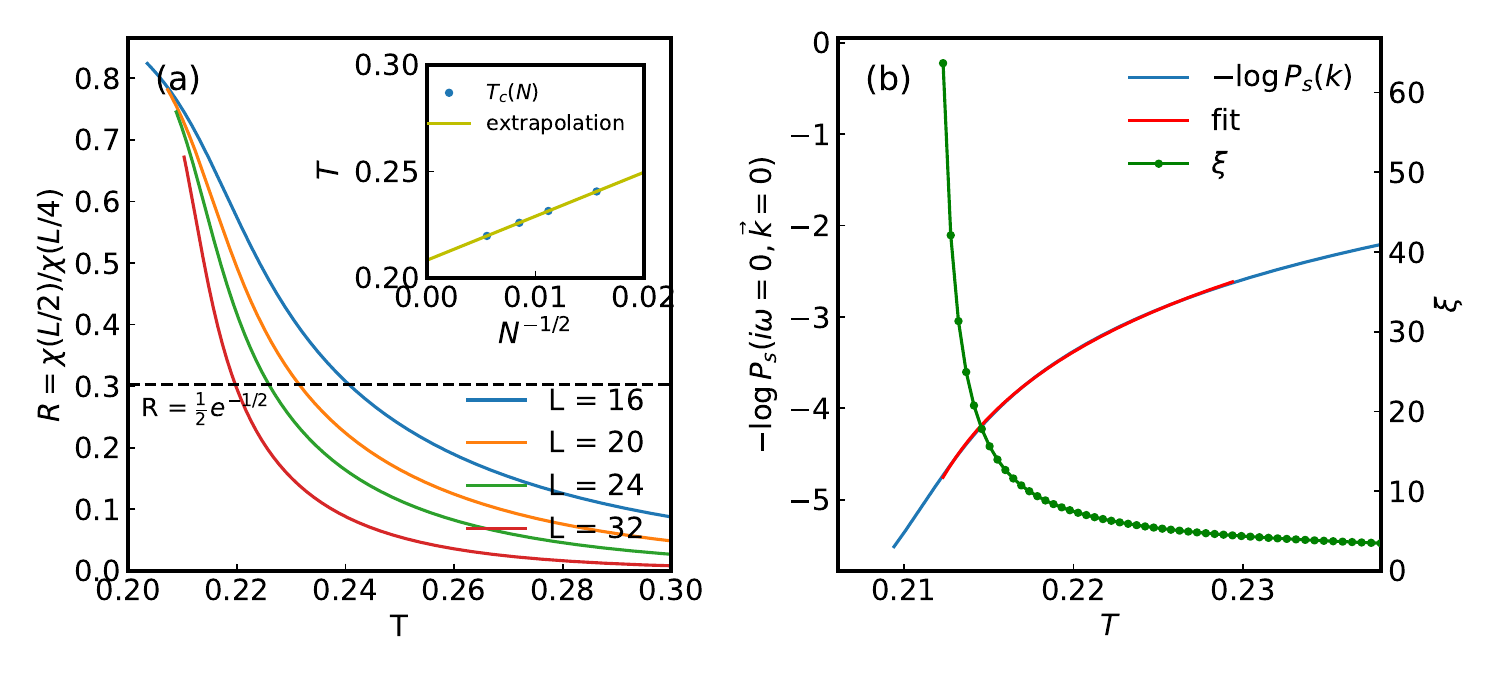}
    \caption{\label{fig:panju} 
    (a) GW results for the s-wave pair correlation in the case 
    $U = -4$ and $\langle n \rangle = 0.5$ for different lattice sizes. 
    The inset shows the critical temperatures corresponding to various 
    lattice sizes, with the yellow line representing an extrapolation 
    fit based on Eq.~(\ref*{tcfss}). The $x$-axis value of zero corresponds to the 
    transition temperature in the thermodynamic limit.
    (b) GW method's critical exponent fitting for the case 
    $U = -4$ and $\langle n \rangle = 0.5$. 
    The blue line represents the logarithmic result for the instability, 
    the green line shows the fitted correlation length, 
    and the red line corresponds to the fitting of the 
    logarithmic instability data within the critical region. }
    \end{figure*}
    \begin{table}[]
        \caption{\label{tab:critical_result}%
                This table presents the results of the critical exponents calculated 
        using the GW method on a $40 \times 40 \times 40$ lattice. 
        The values are derived by fitting the pairing correlation 
        function within different fitting intervals. 
        }
        \begin{ruledtabular}
            \begin{tabular}{p{1cm}|p{1cm}|p{1cm}|p{1cm}|p{1cm}}
                \textrm{ U }&
                \textrm{ n }&
                \textrm{$\gamma$}&
                \textrm{$\nu$}&
                \textrm{$T_c$}\\
                \colrule
                -4 & 0.50 & 1.322 & 0.736 & 0.208 \\
                -4 & 0.50 & 1.310 & 0.671 & 0.208 \\
                -4 & 0.50 & 1.296 & 0.676 & 0.208 \\
                -4 & 0.50 & 1.322 & 0.699 & 0.208 \\
                -4 & 0.50 & 1.297 & 0.688 & 0.208 \\
                -4 & 0.50 & 1.310 & 0.671 & 0.208 \\
                \end{tabular}
        \end{ruledtabular}
    \end{table}
    \begin{table}[]
        \caption{\label{tab:critical_result_HGW}%
        This table presents the results of the critical exponents calculated 
        using the HGW method on a $40 \times 40 \times 40$ lattice. 
        The values are derived by fitting the pairing correlation 
        function within different fitting intervals. 
        }
        \begin{ruledtabular}
            \begin{tabular}{p{1cm}|p{1cm}|p{1cm}|p{1cm}|p{1cm}}
                \textrm{ U }&
                \textrm{ n }&
                \textrm{$\gamma$}&
                \textrm{$\nu$}&
                \textrm{$T_c$}\\
                \colrule
                -8 & 0.50 & 1.904 & 0.878 & 0.270\\
                -8 & 0.50 & 1.873 & 0.851 & 0.271\\
                -8 & 0.50 & 1.873 & 0.920 & 0.271\\
                -8 & 0.50 & 1.904 & 0.948 & 0.270\\
                -8 & 0.50 & 1.873 & 0.996 & 0.271\\
                \end{tabular}
        \end{ruledtabular}
    \end{table}
Here, we determine the phase transition temperature in the thermodynamic limit through 
finite-size scaling analysis of critical behavior. As discussed in the main text, 
the pairing correlation function in finite-size systems does not exhibit true divergence. 
Furthermore, using the divergence of instabilities as a reference can lead to ambiguous 
results due to convergence issues in self-consistent iterations. To address this, 
we establish the phase transition temperature through the following procedure:
First, we calculate the real-space pairing correlation function. For three-dimensional 
systems in the normal phase, the spatial dependence of the correlation function follows 
the behavior described by Eq.~(\ref*{critical length}). 
We determine the $\xi/L$ by comparing the correlation function values at different spatial positions. 
For numerical stability, we specifically compare the values at $L/2$ and $L/4$. 
Based on Eq.~(\ref*{eq:xi}), we identify the critical point when $\xi=L/2$, 
which yields:
$R\equiv\chi(L/2)/\chi(L/4)=\frac{1}{2}e^{-\frac{1}{2}}.$
Our calculations show that the final results are robust against moderate variations in the specific criterion within the same system size scale. 
Finally, we extrapolate to the thermodynamic limit using the finite-size 
scaling relation in Eq.~(\ref*{tcfss}). Fig.~\ref*{fig:panju}(a) illustrates 
this procedure for a representative parameter set, demonstrating 
agreement with the finite-size scaling behavior predicted by Eq.~(\ref*{tcfss}). 
Our finite-size scaling analysis reveals that the extrapolated phase transition temperature in the thermodynamic limit is robust against the precise choice of the correlation length threshold for identifying criticality in finite systems, provided the threshold scale remains commensurate with the system size.
 
To calculate the critical exponent, we first compute the corresponding 
pair correlation function in real space. Using the formula in Eq.~(\ref*{critical length}), 
we fit the correlation function results to obtain the correlation length. 
Since the chosen real-space range affects the calculation of the correlation length, 
we selected multiple fitting ranges in a system of size 40 and 
performed fits over different critical temperature ranges. 
One of the fitting results is shown in Fig.~\ref*{fig:panju}. 
The results from multiple fittings are provided in Table.~\ref*{tab:critical_result} for GW method 
and \ref*{tab:critical_result_HGW} for HGW method. 
By performing statistical analysis on these results, 
we obtain the final values presented in Table.~\ref*{tab:critical}.

\nocite{1}

\bibliography{apssamp}

\end{document}